\documentclass{aa}  

\usepackage{graphicx}
\usepackage{txfonts}
\usepackage{amsmath}
\usepackage{natbib}

\begin{document}

   \title{Size evolution of star-forming galaxies with $2<z<4.5$ in the VIMOS Ultra-Deep Survey \thanks{Based on data obtained with the European 
          Southern Observatory Very Large Telescope, Paranal, Chile, under Large
          Program 185.A--0791. }}

\author{
B. Ribeiro \inst{1}
\and O.~Le F\`evre\inst{1}
\and L. A. M.~Tasca\inst{1}
\and B.~C.~Lemaux \inst{1}
\and P.~Cassata\inst{1,2}
\and B.~Garilli\inst{3}
\and D.~Maccagni\inst{3}
\and G.~Zamorani \inst{4}
\and E.~Zucca\inst{4}
\and R.~Amor\'in\inst{5}
\and S.~Bardelli\inst{4}
\and A.~Fontana\inst{5}
\and M.~Giavalisco\inst{6}
\and N.P.~Hathi\inst{1}
\and A.~Koekemoer\inst{7}
\and J.~Pforr\inst{1}
\and L.~Tresse\inst{1}
\and J.~Dunlop\inst{8}
}

\institute{Aix Marseille Universit\'e, CNRS, LAM (Laboratoire d'Astrophysique de Marseille) UMR 7326, 13388, Marseille, France
\and
Instituto de Fisica y Astronom\'ia, Facultad de Ciencias, Universidad de Valpara\'iso, Gran Breta$\rm{\tilde{n}}$a 1111, Playa Ancha, Valpara\'iso Chile
\and
INAF--IASF Milano, via Bassini 15, I--20133, Milano, Italy
\and
INAF--Osservatorio Astronomico di Bologna, via Ranzani, 1 - 40127, Bologna 
\and
INAF--Osservatorio Astronomico di Roma, via di Frascati 33, I-00040, Monte Porzio Catone, Italy
\and
Astronomy Department, University of Massachusetts, Amherst, MA 01003, USA
\and
Space Telescope Science Institute, 3700 San Martin Drive, Baltimore, MD 21218, USA
\and
SUPA, Institute for Astronomy, University of Edinburgh, Royal Observatory, Edinburgh, EH9 3HJ, United Kingdom
 \\ \\
             \email{bruno.ribeiro@lam.fr}
}

   \date{}

 
  \abstract
   {The size of a galaxy encapsulates the signature of the different physical processes driving its evolution. The distribution of galaxy sizes in the universe as a function of cosmic time is therefore a key to understand galaxy evolution. }
   {We aim to measure the average sizes and size distributions of galaxies as they are assembling before the peak in the comoving star formation rate density of the universe to better understand the evolution of galaxies across cosmic time.}
   {We use a sample of $\sim1200$ galaxies with confirmed spectroscopic redshifts $2 \leq z_{spec} \leq 4.5$ in the VIMOS Ultra Deep Survey (VUDS), representative of star-forming galaxies with $i_\mathrm{AB} \leq 25$. 
We first derive galaxy sizes applying a classical parametric profile fitting method using GALFIT. We then measure the total pixel area covered by a galaxy above a given surface brightness threshold, which overcomes the difficulty of measuring sizes of  galaxies with irregular shapes. We then compare the results obtained for the equivalent circularized radius enclosing 100\% of the measured galaxy light $r_T^{100}$ to those obtained with the effective radius $r_{e,\mathrm{circ}}$ measured with GALFIT.}
   {We find that the sizes of galaxies computed with our non-parametric approach span a large range but remain roughly constant on average with a median value  $r_T^{100}\sim2.2$ kpc for galaxies with $2<z<4.5$. This is in stark contrast with the strong downward evolution of $r_e$ with increasing redshift, down to sizes of $<1$ kpc at $z\sim4.5$. We analyze the difference and find that parametric fitting of complex, asymmetric, multi-component  galaxies is severely underestimating their sizes.  By comparing $r_T^{100}$ with physical parameters obtained through SED fitting we find that the star-forming galaxies that are the largest at any redshift are, on average, more massive and more star-forming. We  discover that galaxies present more concentrated light profiles as we move towards higher redshifts. We interpret these results as the signature of several, possibly different, evolutionary paths of galaxies in their early stages of assembly, including major and minor merging or star-formation in multiple bright regions.}
   {}
   \maketitle
%

\section{Introduction}\label{sec:intro}
Galaxy formation, and early stage evolution, is believed to be a turbulent process where gas inflows, strong winds and galaxy-galaxy interactions give rise to the intricate shapes we encounter in deep, high-$z$, HST observations \citep[e.g.][]{law2007,serrano2010,buitrago2013,mortlock2013,talia2014,guo2015}. A simple, yet fundamental, shape parameter is the galaxy size. This quantity, together with other physical parameters like stellar mass and star-formation rate, is one of the basic ingredients that can help to elaborate a galaxy evolution scenario.

Although it is a simple concept, obtaining galaxy sizes is not an easy task and is subject to a number of assumptions. The most common way to derive galaxy sizes is by performing light profile fitting assuming a given shape for the surface brightness profile using a $\chi^2$ minimization 
\citep[e.g.][]{ravindranath2004,ravindranath2006,trujillo2006a,akiyama2008,franx2008,tasca2009, williams2010,mosleh2011,huang2013,ono2013,stott2013,morishita2014,vanderwel2014,straatman2015,shibuya2015}.  Another  method assumes circular or elliptical apertures around a pre-defined galactic centre and computes the size enclosing a given percentage of the total galaxy flux \citep[e.g.][]{ferguson2004,bouwens2004,hathi2008b,oesch2010,ichikawa2012,curtis-lake2014}. A third approach, involving counting the number of pixels belonging to the galaxy to derive its size, was also explored in \citet{law2007}.

Studies of galaxy sizes at $z>2$ became possible with the deep imaging obtained with the \emph{Hubble Space Telescope}. The first reports on size evolution found that galaxy  sizes as observed in the UV rest-frame were becoming smaller at the highest redshifts \citep{bouwens2003,ferguson2004,bouwens2004}. We have now access to the size evolution up to $z\sim10$ from the deepest HST imaging data \citep[e.g.-][]{hathi2008b,ono2013,kawamata2014,holwerda2014,shibuya2015}. 
With the multi-wavelength and near-infrared coverage of CANDELS \citep{grogin2011,koekemoer2011}  optical rest-frame measurements are reported up to $z\sim3$ for a large collection of galaxies in diverse populations \citep[e.g.][]{vanderwel2014,morishita2014}. 

Despite the existence of morphological differences when galaxies are observed in different bands, size evolution seems to be rather consistent when studied in different rest-frame bands \citep{shibuya2015}. It is generally accepted that galaxy sizes tend to decrease with increasing redshift \citep[e.g.][]{bouwens2003,bouwens2004,ferguson2004,mosleh2012} and that galaxy sizes depend on stellar mass \citep[e.g.][]{franx2008,vanderwel2014,morishita2014} and luminosity \citep[e.g.][]{grazian2012,huang2013}. However some results point to a scenario consistent with no size evolution as seen in UV rest-frame from HST data \citep{law2007,curtis-lake2014} and, at a fixed stellar mass, from optical rest-frame ground-based data \citep{ichikawa2012,stott2013}.

Most samples in the literature that are used to measure sizes at high redshift are composed of galaxies selected through the dropout technique \citep{steidel1999} or photometric redshifts. Typical 1$\sigma$ errors in photometric redshift measurements of dz$\sim$0.2-0.3 add an error of $\sim$5\% (increasing from $z=2$ to $z=4.5$) to angular diameter measurements. This is a small but sizeable error that can be reduced when measuring sizes for galaxy samples with known spectroscopic redshifts. Moreover, having a precise knowledge of the redshift of the galaxy minimizes the ambiguity on which rest-frame bands are observed and allows an accurate estimate of surface brightness dimming effects which are important to reduce measurement scatter for a consistent analysis in a large redshift range. The knowledge of the spectroscopic redshift and the availability of spectra also allow to better characterize the physical properties of the galaxies for which sizes are measured. Uncertainties on stellar masses, star formation rates, and ages, are reduced (R. Thomas et al. submitted) and spectral features like Lyman$-\alpha$ emission can be correlated with galaxy sizes.

In this paper we present the evolution of the size and size distribution of a sample of $\sim$1200 star-forming galaxies with spectroscopic redshifts  $2<z_{spec}<4.5$ selected from the VIMOS Ultra-Deep Survey  \citep[VUDS,][]{lefevre2015}. In addition to using standard profile fitting to derive effective radii, we develop and apply a non-parametric method \citep[adapted from][]{law2007} to compute galaxy sizes which takes into account the surface brightness dimming effect and does not require any assumption on the symmetry of the studied sources. We test the impact of using different rest-frame bands (for a subset of our galaxies) and different methods when deriving size evolution. We then probe the evolution of light profiles of galaxies using image stacks and comparing the concentration of light across the redshift range studied here.

This paper is organized as follows. In section \ref{sec:data_sample} we give an overview of the sample we are working with, how galaxies are selected and describe the imaging data. In sections \ref{sec:re} and \ref{sec:rtot} we describe the methods used to obtain size measurements and their inherent assumptions, and present results on effective radii and non-parametric sizes, respectively. In section \ref{sec:stacks} we present results based on stacked images/profiles. We discuss the redshift evolution of galaxy sizes in section \ref{sec:size_evolution}. 
Section \ref{sec:size_physic} is dedicated to the relation between some physical parameters derived from SED fitting and the sizes of our galaxies.
We discuss and summarize our results in sections \ref{sec:discussion} and \ref{sec:conclusions}. 

We use a cosmology with $H_0=70~km~s^{-1}~Mpc^{-1}$, 
$\Omega_{0,\Lambda}=0.7$ and $\Omega_{0,m}=0.3$. 
All magnitudes are given in the AB system.


\section{Data and Sample Selection}\label{sec:data_sample}

VUDS is a large spectroscopic survey that targeted $\sim10 000$ objects covering an area of 1deg$^{2}$ on three separate fields: COSMOS, ECDFS and VVDS-02h. With the objective to observe galaxies in the redshift range $2<z<6+$, targets were selected based on the first or second photometric redshift peaks being at $z_\mathrm{phot}+1\sigma>2.4$, combined with a colour selection criteria based on the spectral energy distribution (SED) and with flux limits $22.5\leq i_{AB} \leq 25$. To fill in the remaining available space on the masks, a randomly flux selected sample with $23 < i_{AB} < 25$ is added to the target list. The spectra were obtained using the VIMOS spectrograph on the ESO-VLT covering, with two low resolution grisms (R=230), a wavelength range of $3650\AA < \lambda < 9350\AA$. The total integration  is of $\sim 14$h per pointing and grism.

Data processing is performed within the VIPGI environment \citep{scoddeggio2005}, and is followed by extensive redshift measurements campaigns using the EZ redshift measurement engine  \citep{garilli2010}. At the end of this process each galaxy has  flux and wavelength calibrated 2D and 1D spectra, a spectroscopic redshift measurement and associated redshift reliability flag. For more information on this process we refer the reader to \citet{lefevre2015}.


  \subsection{Imaging data in the COSMOS and ECDFS fieldsfield}\label{sec:imaging_data}

To conduct a morphological study of high-redshift galaxies, one should use the deepest, best resolution images coming from  Hubble Space Telescope imaging surveys. Covering the largest contiguous sky region, the COSMOS \citep{scoville2007,koekemoer2007} survey has $\approx2\mathrm{deg^2}$ of ACS F814W coverage down to a limiting magnitude of 27.2 ($5\sigma$ point-source detection limit). As it covers the entirety of the VUDS COSMOS pointings it is the most complete image set to undertake morphological studies with our sample albeit only covering the UV rest-frame.

The more recent CANDELS survey \citep{grogin2011,koekemoer2011} covers five different sky regions, including the COSMOS and ECDFS fields with a large overlap with VUDS fields \citep[see ][]{lefevre2015}. The area covered is smaller than VUDS but near-infrared coverage (namely F125W and F160W) is important because it allows us to probe optical rest-frame morphologies for the majority of our sample in this field. It reaches depths at $5\sigma$ of 28.4,27.0 and 26.9 at F814W, F125W and F160W respectively, deeper in the F814W band than the aforementioned COSMOS observations. The typical spatial resolution of these images ranges from $0.09"$ ($0.6,0.8$ kpc, at $z=4.5,2.0$) for the F814W band up to $0.18"$ ($1.2,1.5$ kpc, at $z=4.5,2.0$) in the F160W band. The pixel scale of the mosaics used in this  paper are of $0.03''/\mathrm{pixel}$ for F814W images and $0.06''/\mathrm{pixel}$ for F125W/F160W images.

As the multi-wavelength coverage with HST of the large area covered by VUDS  is scarce, we also use deep ground based imaging covering the entire VUDS area on the COSMOS field for a control check on the wavelength dependence of our radii measurements. From the CFHT Legacy Survey (CFHTLS) data release 7\footnotemark{}\footnotetext{\url{http://www.cfht.hawaii.edu/Science/CFHTLS/}} we use the deep i' and z' band imaging  which reach an 80\% completeness limit in AB of 25.4 and 25.0 for point sources, respectively. Extensive near infrared ground-based imaging is also available on the COSMOS field. 
The UltraVISTA survey covers 1.5 square degree of the field and provides the deepest observations in the near-infrared bands. The $Y,J,H$ and $K_s$ UltraVISTA DR2 images reach $5\sigma$ limiting magnitudes of 24.8 (25.4), 24.6 (25.1), 24.7 (24.7) and 23.98 (24.8) AB in the deep (ultra-deep) regions \citep{mccracken2012}.

In addition to the CANDELS observation in ECDFS, there are also publicly available F850LP observations from GEMS \citep{rix2004}. However, the VUDS pointings were designed to maximize the overlap with the CANDELS area in ECDFS, thus, adding GEMS observations cause only a small increase in the number of sources in the sample (68 additional sources to the stellar mass selected sample defined in the next section). For that reason we exclude data from these surveys from the size analysis presented in this paper. The GOODS survey \citep{giavalisco2004} covers a similar area as the CANDELS observations and thus we have opted to not include images in our analysis.


  \subsection{A stellar mass-selected sub-sample from the VUDS spectroscopic survey}\label{sec:mass_selection}
  
With the knowledge of the spectroscopic redshift, SED fitting was performed on the VUDS sample using Le Phare \citep{arnouts1999,ilbert2006} applied to the extensive photometric data available in the  COSMOS and ECDFS fields as described in Sect. \ref{sec:imaging_data}.

The SED fitting procedure closely follows the method described in \citet{ilbert2013}, and the specifics for VUDS are detailed in \citet{tasca2015}. The two main parameters of interest in this paper are the stellar mass, M$_{\star}$, and the star formation rate ($SFR$) for which the median values of the probability density function are used. We refer the reader to \citet{ilbert2013} for typical uncertainties on these quantities \citep[see also][]{tasca2015}. We also use the physical parameters derived from the simultaneous SED fitting of the VUDS spectra and all multi-wavelength photometry available for each galaxy, using the code GOSSIP+ as described in Thomas et al. (submitted). This method expands the now classical SED fitting technique to the use of UV rest-frame spectra in addition to photometry, further improving the accuracy of key physical parameter measurements (see Thomas et al. for details).  It also provides measurements of physical quantities such as galaxy ages as well as the IGM transmission along the line of sight of each galaxy, an improvement compared to using a fixed transmission at a given redshift \citep{thomas2015}. 
The sample selection (discussed below) and color correction detailed in section \ref{sec:rtot_color} rely on the parametered derived from LePhare. The comparison of sizes with physical parameters descrbied in section \ref{sec:size_physic} uses the results from GOSSIP+. We stress that the results presented in this paper are robust against the method used to derive the stellar mass selection of our sample.

\begin{figure}
   \centering
   \includegraphics[width=\linewidth]{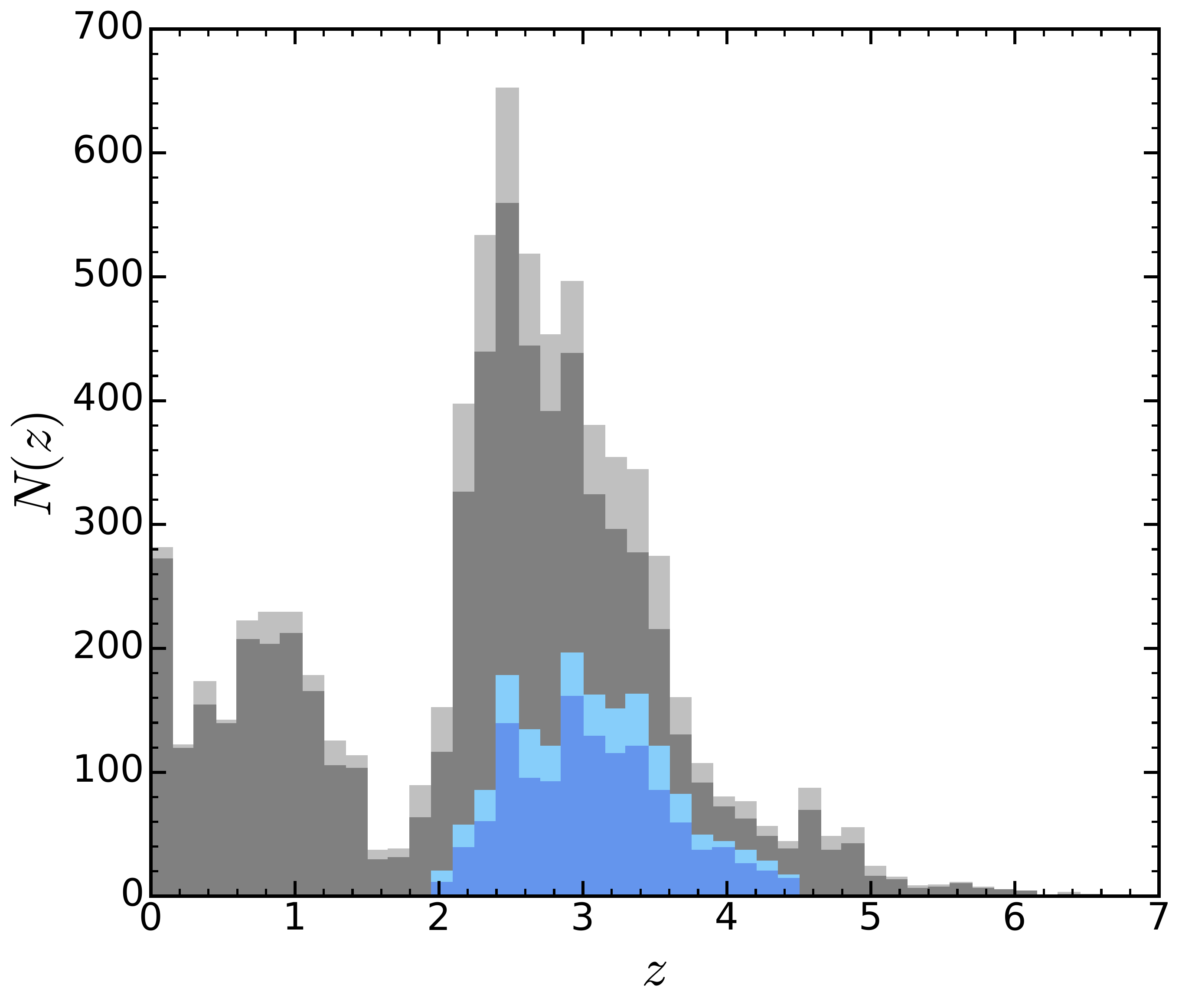}
      \caption{The N(z) for the entire VUDS sample considering all galaxies with a redshift measurement in light grey and those with a redshift reliability flag corresponding to a $>75\%$ certainty in dark grey. In blue colors we present the distribution for the COSMOS+ECDFS, stellar mass selected and $2<z<4.5$ sample (all flags in light blue and reliable flags in darker blue).}
         \label{fig:zed_distribution}
   \end{figure}

To follow the evolution of galaxy sizes in a similar population as a function of redshift, we define our sample imposing an evolving lower stellar mass limit. This choice implies that stellar masses of galaxy populations at different redshifts follow the general stellar mass growth of star-forming galaxies, broadly representing the same coeval population. With this simple evolving stellar mass cut we could possibly miss galaxies if their properties make them fall out from the VUDS selection function, i.e. if galaxies selected at the high redshift end of our sample become quiescent by the time they reach $z\sim2$. The average size we report would then be biased against these galaxies at the lower redshift end of our survey. The other possible choice that could be made is to opt for a constant stellar mass limit at all redshifts. In this condition, the sample would contain more lower stellar mass star-forming galaxies at lower redshifts. If the stellar mass-size relation is not evolving with redshift \citep[see e.g.][]{vanderwel2014,morishita2014} then this would add galaxies with lower sizes in the lower redshift bins of our study, complicating the comparisons of size distributions.  We define a lower stellar mass limit in our sample anchored at $z=4.5$ as $\log_{10}(M_\star/\mathrm{M}_\odot)>9.35$ (below which the VUDS sampling drops). We then use the stellar mass function evolution from \citet{ilbert2013} together with the typical sSFR of VUDS galaxies \citep{tasca2015} to follow the typical stellar mass growth of VUDS galaxies, and define the stellar mass selection threshold at different redshifts using
\begin{equation}
\log_{10}(M_\star/\mathrm{M}_\odot) > -0.204(z-4.5)+9.35.
\label{eq:mass_sel}
\end{equation}
In figure \ref{fig:zed_distribution} one can see the redshift distribution of the entire VUDS sample and that of the stellar mass-selected sample in the redshift range considered defined for the purpose of this study. Such selection translates into 1645 galaxies in the COSMOS and ECDFS fields with a spectroscopic redshift measurements and with stellar masses obeying equation \ref{eq:mass_sel} of which 1242 have spectroscopic redshift flags that have a reliability of $>75\%$ \citep[flags X2,X3,X4 and X9 with X=0,1,2 see][for more details]{lefevre2015}.

\begin{figure*}
   \centering
   \includegraphics[width=\linewidth]{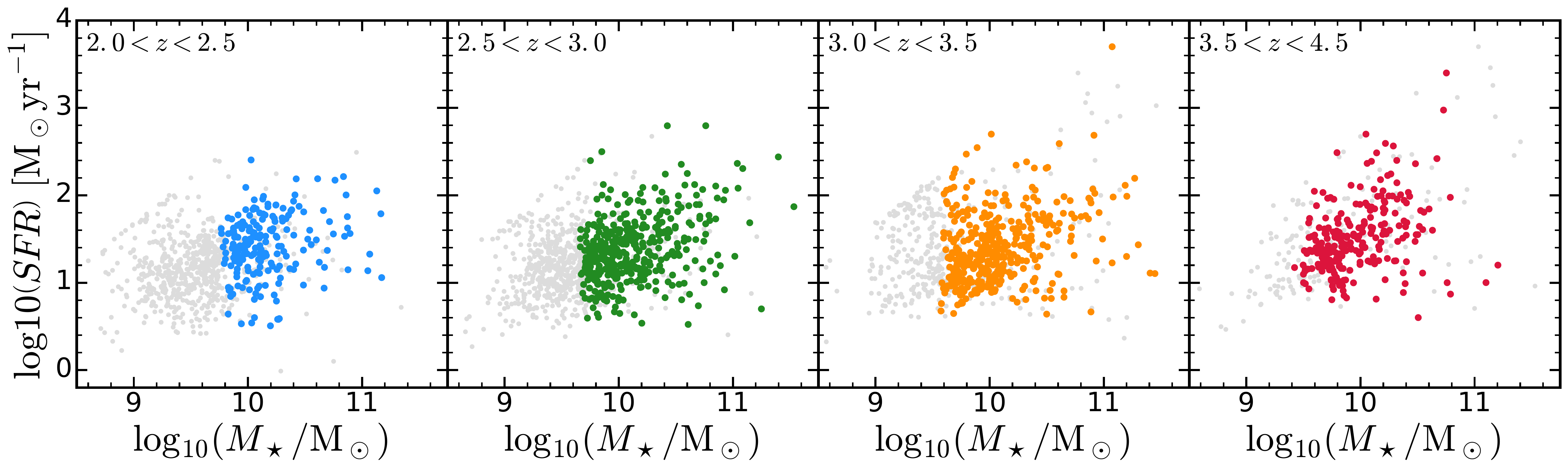}
      \caption{Range in stellar masses and SFRs for the sample studied here. In each panel, the light gray points refer to the entire VUDS sample in each redshift bin. The colored points are the galaxies in the selected stellar mass range and with reliable spectroscopic flags (correct redshift probability of $>75\%$, see text for more details).}
         \label{fig:mass_sfr_plot}
   \end{figure*}

\begin{figure*}
   \centering
   \includegraphics[width=\linewidth]{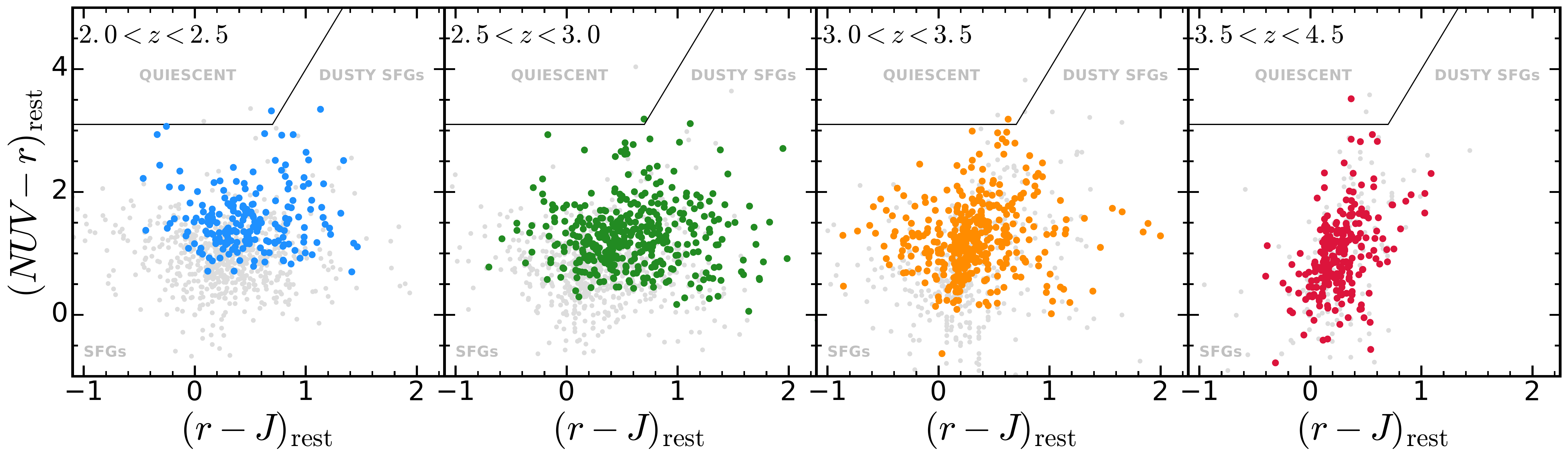}
      \caption{NUVrJ diagram \citep{ilbert2010,ilbert2013} for the sample studied here in different redshift bins. In each panel the colors have the same meaning as in figure \ref{fig:mass_sfr_plot}.}
         \label{fig:NUVrJ_diagram}
   \end{figure*}

We show the stellar mass-SFR relation of our sample divided in four redshift bins in figure \ref{fig:mass_sfr_plot}. It shows that our sample probes galaxies with typical median stellar masses of $10^{10}\mathrm{M_\odot}$ and ranging from $\log_{10}(M_\star/M_\odot)\gtrsim 9.35$ and up to $10^{11}\mathrm{M_\odot}$. In terms of SFR, our galaxies are in the range $0.5\lesssim \log_{10}(SFR)\lesssim 3.0$ at all redshifts with median values of $\log_{10}(SFR)\sim 1.4$.
To further characterize our galaxy population we plot in figure \ref{fig:NUVrJ_diagram} the position of each galaxy with respect to the quiescent region as defined in \citet{ilbert2013}. The lack of galaxies in the upper right part of the panels in the NUVrJ diagram indicates that we are missing dusty star-forming galaxies. We also lack the quiescent population (only 4 galaxies fall in the quiescent region of the diagram). This confirms that our sample is probing the bulk of the massive, low dust star-forming population at the redshifts considered. 

  
\section{Parametric size measurements: Effective radius $r_e$}\label{sec:re}
One of the most widespread ways of measuring galaxy sizes at all redshifts is by parameterizing their surface brightness profiles with a given functional form \citep[e.g.][]{ravindranath2004,ravindranath2006,trujillo2006a,akiyama2008,franx2008,williams2010,mosleh2011,huang2013,morishita2014,vanderwel2014,straatman2015,shibuya2015}. The \citet{sersic1968} profile is the preferred choice for fitting galaxy photometry and it is given by
\begin{equation}
\centering
I (r) = I_e \exp[-\kappa(r/r_{e})^{1/n}+\kappa]
\label{eq:sersic}
\end{equation}
where the S\'ersic index $n$ describes the shape of the light profile, $r_e$ is the radius enclosing 50\% of of the total flux, $I_e$ is the surface brightness at radius $r=r_e$ and $\kappa$ is a parameter coupled to $n$ (see for example \citealt{ciotti1999}) such that half of the total flux is enclosed within $r_e$. An index of $n=1$ corresponds to a typical pure disk galaxy, whereas $n=4$ corresponds to the de Vaucouleurs profile which is associated to elliptical galaxies. On 2D images, each S\'ersic model has potentially seven free parameters: the position of the center, given by $x_c$ and $y_c$, the total magnitude of the model, $m_{tot}$, the effective radius, $r_e$, the S\'ersic index, $n$, the axis ratio of the ellipse, $b/a$ and the position angle, $\theta_{PA}$, which refers to the angle between the major axis of the ellipse and the vertical axis and has the sole purpose of rotating the model to match the galaxy's image.


  \subsection{Method}\label{sec:re_method}

To measure the effective radius of our objects we use the 2D surface brightness fitting tool GALFIT \citep{peng2002,peng2010}. For each galaxy we choose to fit a single S\'ersic profile with no constraint in the parameters. To remove the effect of the ellipticity of each source, the final size value as measured from GALFIT is computed as the circularized radius via
\begin{equation}
r_{e,circ} = r_e \sqrt{q},
\end{equation}
where $q$ is the axis ratio ($b/a$) of the elliptical isophotes that best fit the galaxy. In this sense, a comparison with $r_{T}$ defined in section \ref{sec:rtot} is also more immediate, as both are circularized forms of size measurements. The initial parameters were retrieved from running SExtractor \citep{bertin1996}.

  \subsection{PSF and masks}\label{sec:re_psfs}

In order to provide accurate size measurements it is crucial to have a good image representing the point spread function (PSF) of the data. For that purpose a list of individual, bright, non-saturated, field stars were pre-selected based on their position in the stellar locus of the $\mu_\mathrm{max}-mag$ diagram. This procedure was applied to all the CANDELS, UltraVISTA and CFHTLS mosaic images. Then, we selected by visual inspection the stars to be used for the stack. Thus, we built a high S/N image of the PSF for each image by stacking $\sim40$ stars in each CANDELS field. For COSMOS we have used the models built with \emph{TinyTim} \footnotemark{}\footnotetext{\url{http://www.stsci.edu/hst/observatory/focus/TinyTim}} described in \citet{rhodes2007}.

For space based imaging, we fed GALFIT with $6\arcsec \times6\arcsec$ image cutouts around the galaxy. As for the masking procedure, we used the segmentation map produced by SExtractor and flagged all companion objects for which all pixels associated with it are at a distance greater than 1\arcsec of the VUDS target. The flagged pixels are not taken into account when doing the $\chi^{2}$ fitting by GALFIT. 
Based on deep galaxy counts \citep{capak2007} we estimate the number of companions at a distance smaller than 1\arcsec and brighter than $i_\mathrm{AB}=25.5$  is 6\%.
This reassures us that we are getting size measurements for sources that fall within the slits, i.e. it is unlikely that GALFIT will lock on a companion object.

        \subsection{Image simulations}\label{sec:re_simul}

Simulations of 15\,000 galaxies are performed to test the reliability of the obtained values for the effective radius of the galaxies in the sample. To do so, we use the computed SExtractor source catalog to select 200$\times$200 pixel sky regions on which the simulated galaxies are dropped. The sky regions are randomly selected so that we do not introduce artificial bias by using hand picked regions. However, since the selection of pure sky regions in a random way is very time consuming a method is devised to select the required size regions in a faster way while still taking cautions not to overlap simulated and real sources and to avoid bright sources within the region. This method selects random regions where: there are no pixel detections within a 50 pixel radius from the center of the sky region; the neighboring sources allowed in the stamp image must not be brighter than 20 mag; the number of SExtractor detected sources within the stamp is limited to 10; all pixels have a non-zero value. Then the simulated galaxies have S\'ersic profiles that are randomly generated from a set of values drawn from uniform distributions in the intervals defined in table \ref{tab:simvals}. The profiles are subsequently convolved with the PSF and dropped on extracted regions. Finally all the simulated objects are fed onto the same pipeline as the real images to get their output parameters from GALFIT. The results of these simulations (see figure \ref{fig:simulated_re}) show that one can retrieve the input effective radius value within 10-20\% for 68\% of the simulated galaxies that have a total magnitude $\lesssim25$. There is also a slight trend with the input effective radii in the sense that galaxies with higher input $r_e$ tend to have slightly higher dispersion (up to $\sim20-30\%$) of the difference one gets. We note however that 93\% of the galaxies in our sample have measured radii smaller than 20 pixels for which we obtain a radius within $\sim15\%$ of the input value. We have tested this simulation by using different background images from which to extract the sky regions (COSMOS F814W tiles and CANDELS F814W, F125W, F160W mosaics) and found no significant differences in the simulation results. Similar conclusions were found by, e.g., \citet{mosleh2011,morishita2014}.

\begin{table}
\centering
\caption{Interval of simulated values of artificial galaxies.}
\begin{tabular}{ccccc }
\hline
mag & $r_e$[pixel] & $n$ & $q$ & $\theta_{PA}$ [degrees]\\
\hline
22 - 27 & 0.5 - 30 & 0.5 - 10 & 0.1 - 1.0& -90 - 90\\
\hline
\end{tabular}\label{tab:simvals}
\end{table}

An additional test is carried out to test the robustness of GALFIT results against the choice of first guess parameters. In this case, for each galaxy, we run GALFIT fifty times, each of which with a set of Gaussian randomly deviated values centered on the SExtractor input guesses and with widths that include  at least a 50\% deviation from the input value within 1$\sigma$. We then compare the median value of the fifty runs with the results from the single run with SExtractor first guesses and find that no more than $\sim9\%$ of the galaxies for which GALFIT converged have median values offset by more than $3\sigma$ and that most of the galaxies ($\sim 90\%$) are within $1\sigma$. The outliers tend to happen for cases where GALFIT does not converge for one or more of the fifty tries. Nonetheless, the majority of the galaxies for which we have GALFIT results (around $\sim74\%$) allow the convergence more than 40 (out of 50) times.


        \subsection{Dependency with colour}\label{sec:re_color}

 \begin{figure}
\centering
\includegraphics[width=\linewidth]{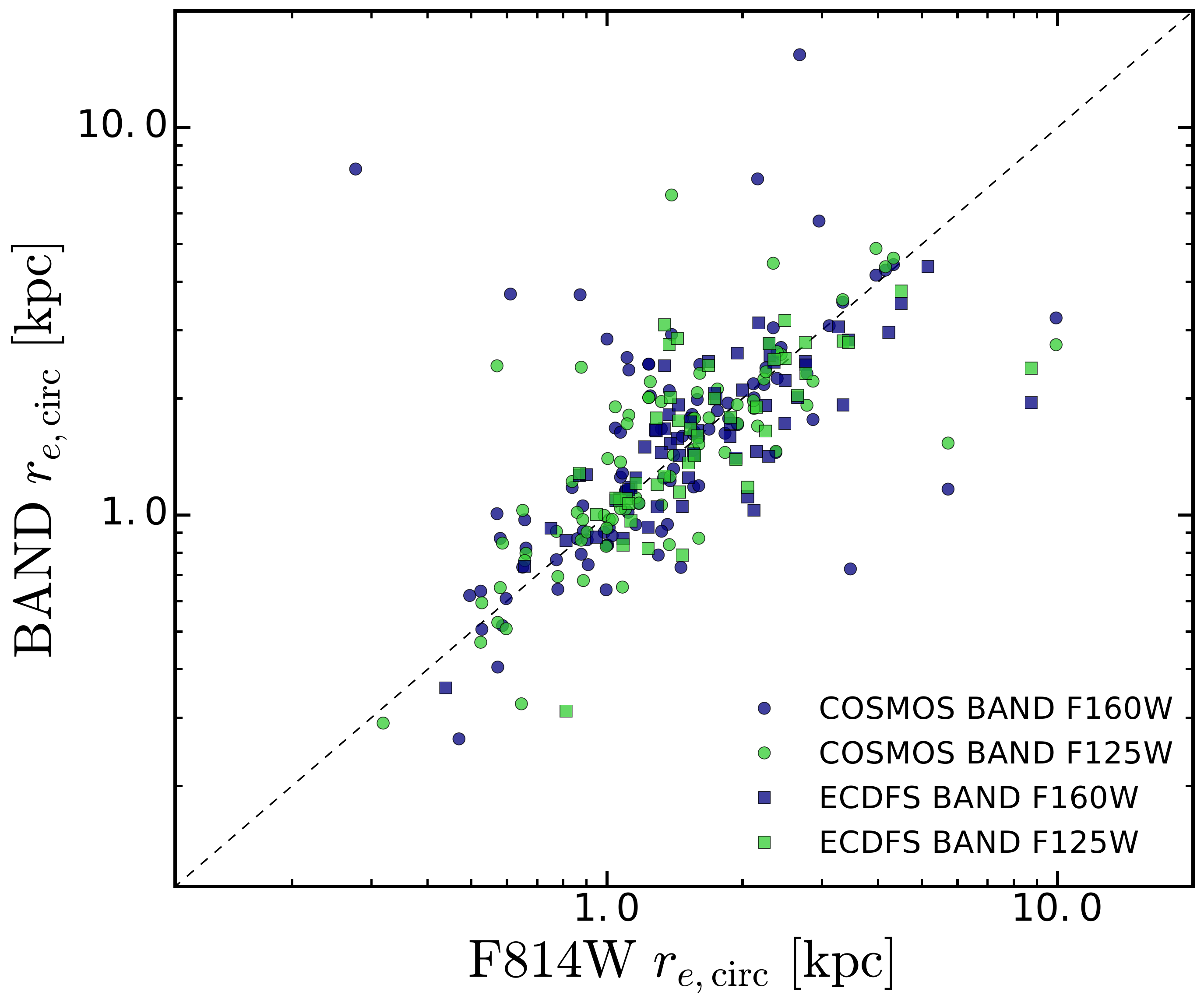}
\caption{Comparison of size measurements derived with GALFIT for a subset of 153 (118) VUDS galaxies at $2<z<4.5$ with available F814W and F160W (F125W) images from CANDELS. Circles and squares refer to galaxies from COSMOS and ECDFS, respectively. The dashed black line is showing where both radii are equal.}
\label{fig:color_comparison_re}
\end{figure}

To study size evolution across a large redshift range one should use common rest-frame measurements whenever possible. When not possible some approximations must be made. The simplest one consists on using the observed band closest to the rest-frame one wants to consider \citep[e.g.][]{morishita2014,shibuya2015}. A more evolved approach is presented by \citet{vanderwel2014} where a wavelength dependence of the measured radii is observed and fitted. A correction is then applied to each galaxy to obtain the value of the radius at exactly the wavelength one chooses. However, such approaches require a multi-wavelength coverage of the targets which is not available for the majority of our sample.

\begin{figure*}
   \centering
   \includegraphics[width=\linewidth]{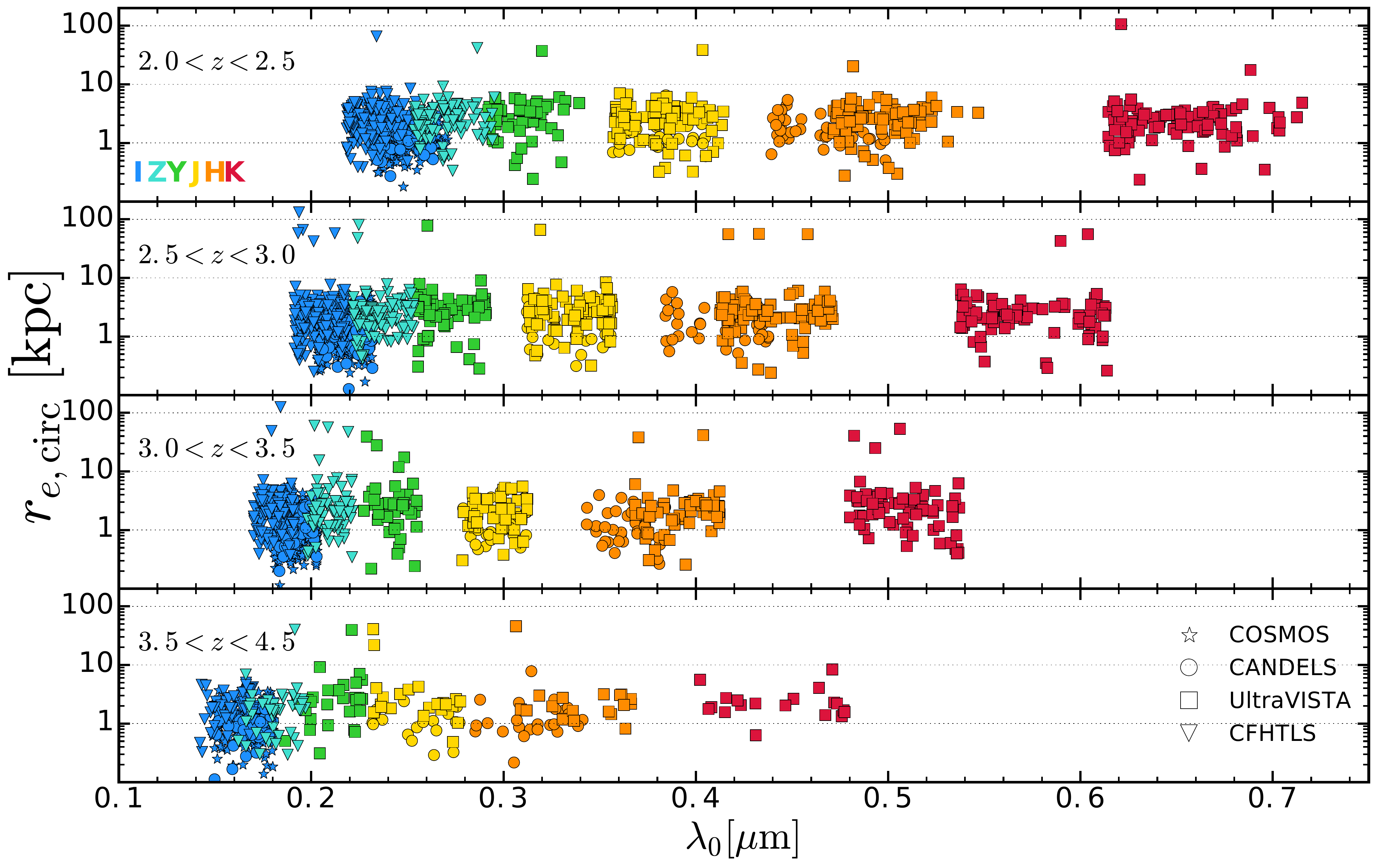}
      \caption{Measured Sizes as a function of wavelength for galaxies with $\log_{10}(M_\star)>9.5$ at $2<z<4.5$ in the COSMOS field. Different symbols correspond to different surveys (symbol coding on the bottom panel) and different colors to different observed bands (color coding in the top panel). Each point represents the size of a given galaxy at the rest-frame wavelength of the pivot of the filter.}
         \label{fig:color_depend}
   \end{figure*}

For those galaxies which are in the CANDELS area (roughly 10\% of our sample), we did compute their effective radii in three different bands: F814W, F125W and F160W. Doing a simple inter-wavelength comparison one finds that the derived sizes are similar in all three bands (see Figure \ref{fig:color_comparison_re}). We report a median offset of 0.04 (0.01) kpc for sizes measured in F160W (F125W). The dispersion on the size measurements across different bands is 0.22 dex and 0.18 dex for F814W-F160W and F814W-F125W comparisons respectively. 
For further inspection, we used the wealth of multi-wavelength observations on the COSMOS field (space and ground based imaging) with a variety of instruments from three different surveys (CFHTLS in the optical, and WIRDS and UltraVISTA in the near-infrared) to obtain size measurements up to the Ks band which, at the highest redshifts considered here, still probes the regions above the Balmer/D$_n$4000 break. Then, for each galaxy in our sample, we plot its size value as a function of the observed wavelength computed using the knowledge of the spectroscopic redshift and the filter wavelength, $\lambda_F$.
\begin{equation}
\lambda_0 = \frac{\lambda_F}{1+z}
\end{equation}
Size measurement results at different wavelengths obtained from our simulations are presented in figure \ref{fig:color_depend}. The overall trend is that, at the redshifts considered here, measurements of circularized effective radii are independent on wavelength to first order.  This result tells us that circularized effective radii computed from F814W observations are not much affected by the rest-frame wavelength at which the galaxies are observed. To quantify the wavelength dependence we have opted to fit the relation
\begin{equation}
\log(r) = \beta_\lambda +\alpha_\lambda \log(\lambda_0)
\label{eq:wave_dep}
\end{equation}
to all the data points considering space based imaging and ground based imaging separately. We find that for ground based data (covering from I to Ks bands) the slopes measured are all consistent with no wavelength dependence ($\alpha_\lambda=0$ is within 1$\sigma$ at all redshift intervals). For space based data (covering from I to H bands) we find slightly positive slopes indicating slightly larger sizes in the redder bands, but only marginally significant 1.9$\sigma$ away from a flat slope $\alpha_\lambda=0$. Using our fit values in equation  \ref{eq:wave_dep} we estimate that the ratio ranges within $r_{1700\AA}/r_{2500\AA}=0.89-1.01$.

In a complemtary test, we fit the size-wavelength relation for all individual galaxies which have measurements in three or more bands. The median slope is $\alpha_\lambda = 0.03\pm0.27$ and $\alpha_\lambda=-0.12\pm0.34$ for space and ground based observations, respectively. This translates into a ratio $r_{1700\AA}/r_{2500\AA}=0.99(1.05)$ for space (ground) based data. This supports the fact that the wavelength dependence of galaxies in our sample is very shallow and consistent with no dependence at all for the ensemble of the galaxies in our sample. We note however that there is a large dispersion amongst our galaxies, which means that the wavelength dependence requires a different treatment when considering a case by case basis.

We have shown in figure \ref{fig:color_depend} that we can measure sizes from ground-based data for the galaxies in our sample. This would allow for the use of the entire VUDS sample, including galaxies in the VVDS-2h field. However, since the convergence success of GALFIT is much less ($\sim30\%$) and the fact that the method described in section \ref{sec:rtot} is PSF dominated preventing us from getting galaxy sizes (sizes are systematically overstimated when compared to those obateined with HST images) we opt to conduct our analysis on space-based data only.
 


\subsection{Results: $r_e$ measurements}\label{sec:re_results}

We fit all galaxies in the stellar mass selected sample using the automated GALFIT procedure described in the previous sections to obtain effective size measurements $r_e$. After fitting all galaxies, a subset of $19.5\%$ of those GALFIT failed to converge. These galaxies are analyzed on an object-per-object basis and a manual re-fitting is attempted by tuning the initial set of parameters. After this manual try, of the 1242 galaxies in our stellar mass selected sample, 263 (15.7\%) objects have no structural parameters because GALFIT failed to converge. These objects are usually very low surface brightness or highly concentrated galaxies.

\begin{figure}
   \centering
   \includegraphics[width=0.49\textwidth]{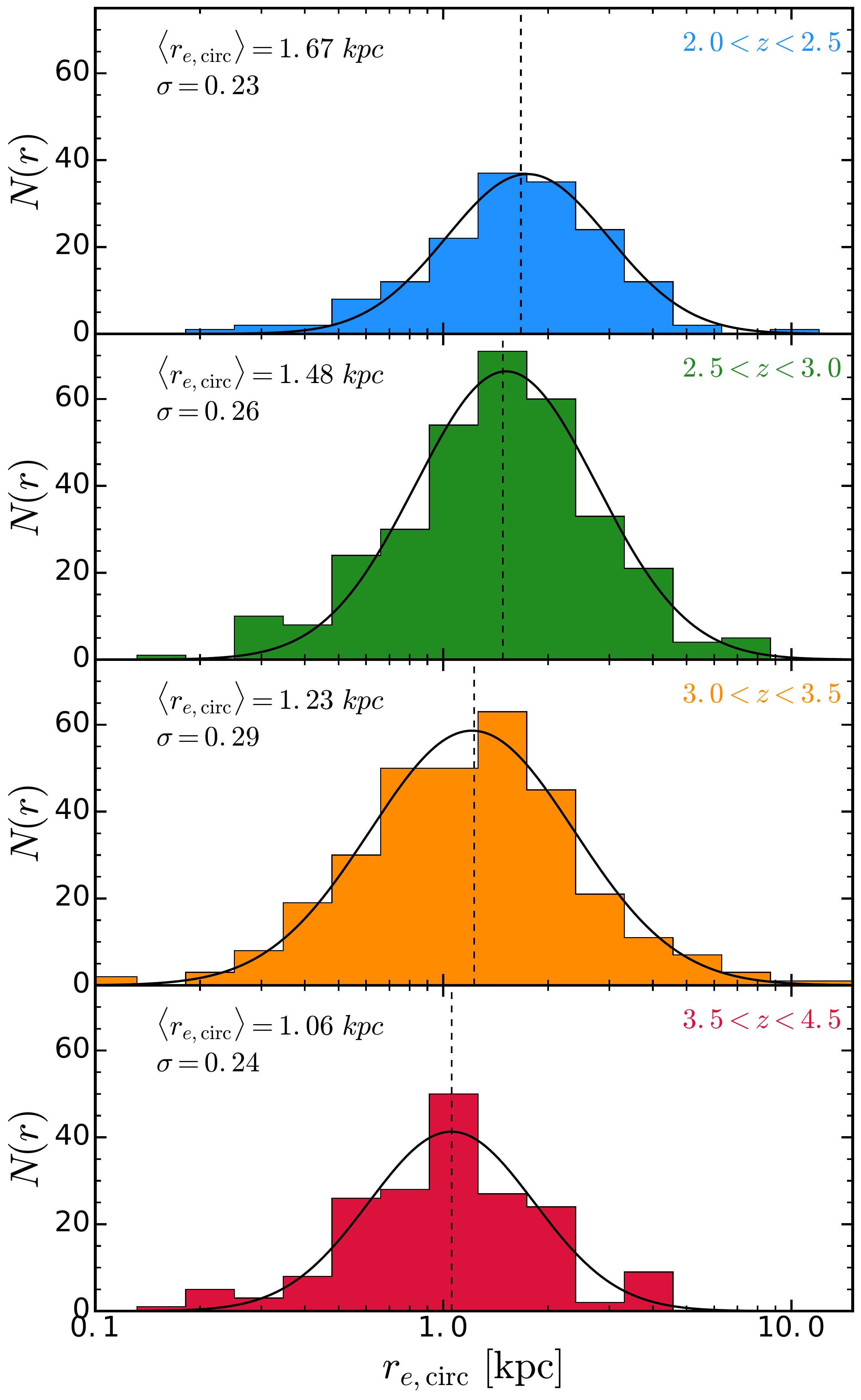}
      \caption{Size distribution for different redshift bins. The vertical dashed line indicates the median value for each bin. The black line is the fit of the distribution using a normal function in log space.}
         \label{fig:re_distribution}
   \end{figure}

The size distributions presented in Figure \ref{fig:re_distribution} show similar shapes at all redshifts showing a slow monotonic decrease in their median sizes as redshift increases. Each distribution is fitted with a normal function (in log space) and the derived sigma values from the fit are similar at all redshifts (0.23,0.26,0.29 and 0.24 with growing redshift).

\begin{figure}
   \centering
   \includegraphics[width=\linewidth]{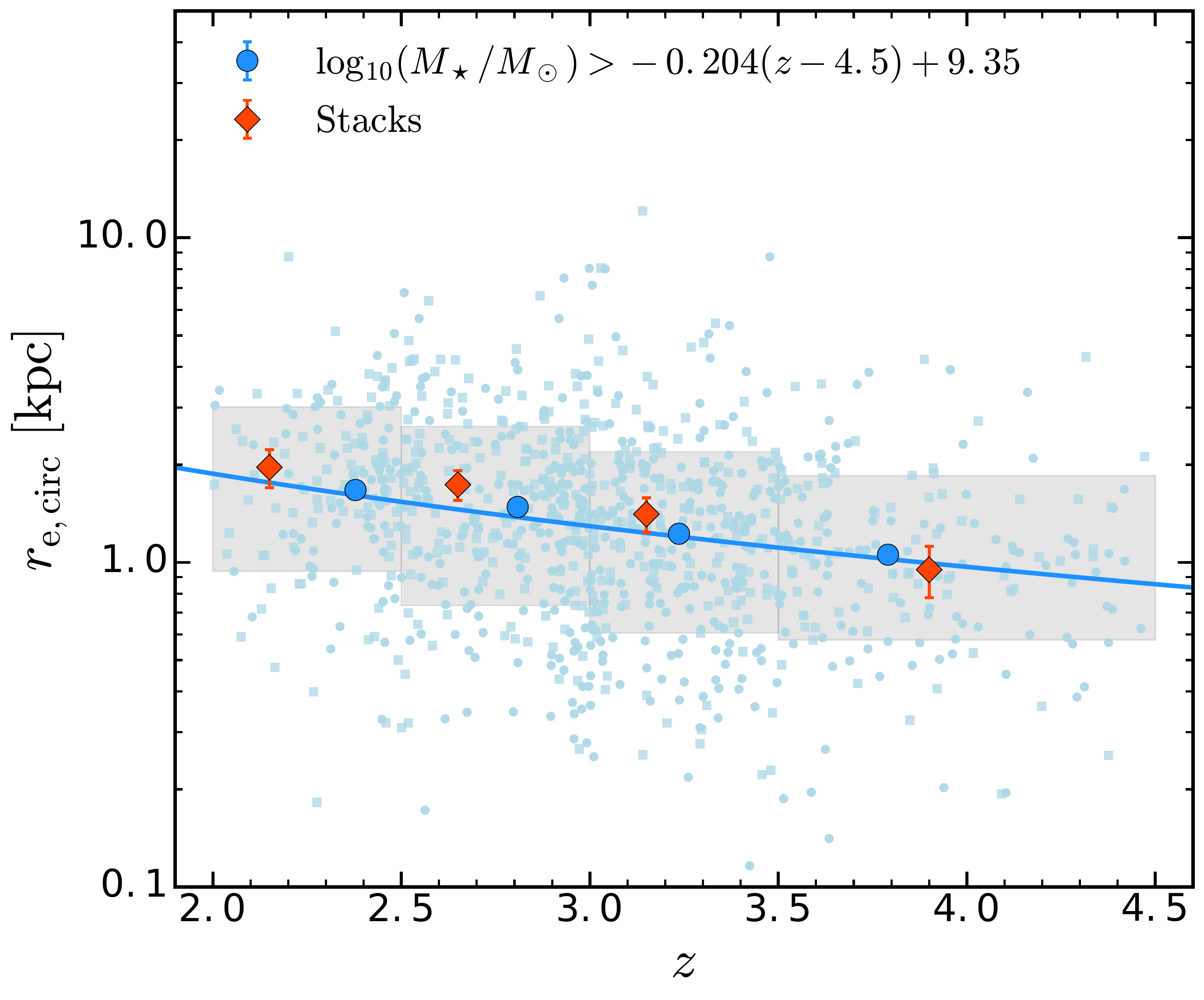}
      \caption{Size evolution with redshift. Each galaxy with a good size measurement is plotted with a small blue point (squares for redshift confidence 2 and 9, circles for 3 and 4). The median values (in redshift and size for each bin) and respective error ($\sigma/\sqrt{N}$) per redshift bin are shown by the large blue points with the error bars. The shaded region delimits the 16th and 84th percentiles including 68\% of the sample in each redshift bin. The red diamonds are the effective radii of stacked images computed from the method described in section \ref{sec:stacks} and are plotted at the center of the redshift bin.}
         \label{fig:re_evolution}
   \end{figure}

The overall evolution of sizes across the redshifts considered here is shown in figure \ref{fig:re_evolution}.
We find that the effective radius of galaxies continuously decreases over the redshift range of our sample. Galaxies with $2< z < 2.5$ have $\langle r_{e,\mathrm{circ}} \rangle=1.67\pm 0.09$ kpc, while in $4 < z < 4.5$ the median effective radius is a factor of 1.6 smaller, $\langle r_{e,\mathrm{circ}} \rangle=1.05\pm 0.05$ kpc. In every redshift bin a large dispersion in the effective radii is measured amongst the galaxy population and for all redshift bins $\sim$9\% of the population has measured radii which are large ($r_{e,circ}>3$ kpc) and some extremely so ($7-10$ kpc) in all redshift bins save the last. These features of the distribution are discussed together with other measurements in Sections \ref{sec:size_evolution} and \ref{sec:conclusions}.

  
\subsection{The impact of cosmological dimming on $r_e$}

A key point to take into account in this analysis is that GALFIT-based measurements do not take into account the surface brightness (SB) dimming effect which is a strong function of redshift. Parametric modeling works well even under low S/N conditions \footnotemark{}\footnotetext{See \url{http://users.obs.carnegiescience.edu/peng/work/galfit/TFAQ.html\#size_and_noise}}, recovering the right parameter values, although with larger uncertainties which should, in principle, deal with the increased SB dimming of galaxies at higher redshift. The dimming effect, if significant, would act in the sense of getting smaller sizes at higher redshifts that would in turn impact the derived evolution into having a steeper slope. Thus, a correction for the cosmological dimming could lead to a weaker size evolution. 

We have tested this scenario by masking the pixels that are below the surface brightness threshold defined in equation \ref{eq:thresh_color} and then run GALFIT to obtain the size measurements from this pixel set reduced to the brightest regions of galaxies. We find that GALFIT finds similar sizes for both cases with and without masking those low surface brightness pixels and that the trend in evolution is not affected. 


\section{Galaxy area and equivalent radius $r_{T}$}\label{sec:rtot}

\begin{figure*}
\centering
\includegraphics[height=0.95\textheight]{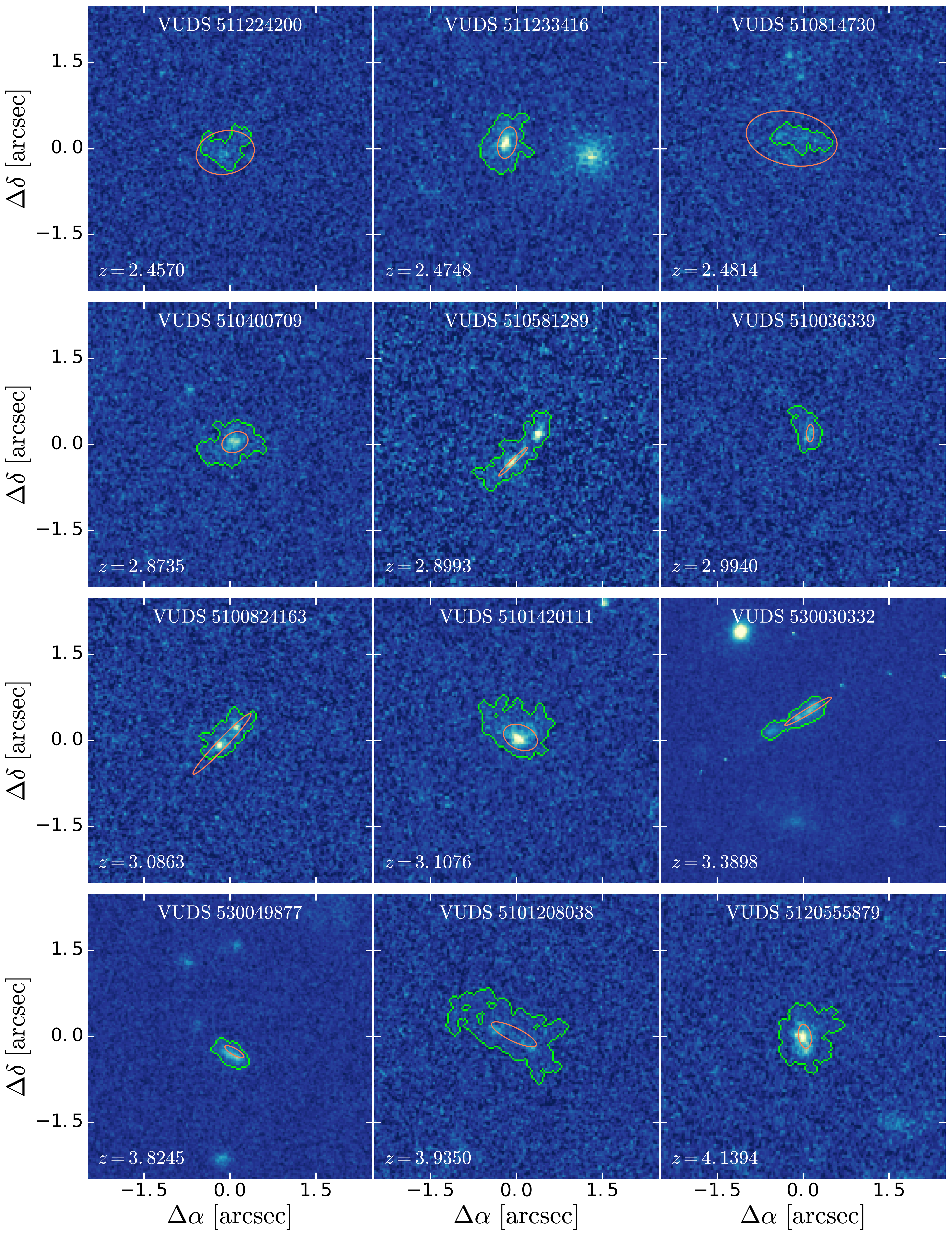}
\caption{F814W imaging in COSMOS \citep{koekemoer2007} and CANDELS \citep[when available,][]{koekemoer2011} of the measured sizes for three galaxies in each redshift bin of our sample. The green contours show the segmentation map of the galaxy at the defined threshold which translates to $T_{100}$ from equation \ref{eq:npsize} measured at $k_p=1.0$. The red ellipse are constructed from the values of $r_e,\ b/a$ and $\theta_\mathrm{PA}$ derived from GALFIT for those galaxies.
}
 \label{fig:clumpy_galaxies}
\end{figure*}

While the local universe is dominated by galaxies with symmetric shape in the form of discs and bulges, the distribution of galaxy shapes becomes more complex as redshift increases. By redshift $z\sim1$ irregular galaxies make up to 52\% of the population of galaxies \citep{serrano2010}. By redshift $z\sim3$ galaxies with irregular shapes dominate  as they represent $\sim 65 (40)\%$ of the population of  massive, $\log_{10}(M_\star)>10 (11)$, galaxies \citep{mortlock2013,buitrago2013}. These irregular shapes often come in the form of clumpy galaxies e.g. representing $\sim 60\%$ of the star-forming galaxies at $2<z<3$ \citep{guo2015}.  
{
This fraction is expected to be even higher when going to higher redshifts \citep[]{conselice2009}.
}
  Irregular shapes can take very diverse forms, including objects with one single asymmetric component, with multiple components, with a tadpole shape, extended with low surface brightness, and can be confused in the presence of close projected or physical companions.  Some examples of galaxies in our sample are shown in Figure \ref{fig:clumpy_galaxies}. The fit with GALFIT gives an effective radius which highly depends on the relative brightness of each clump of the galaxy, but the residuals after fitting show how difficult it is to fit such an object with a symmetric profile.

These considerations prompted us to use other ways to derive sizes of galaxies than parametric fitting. A non-parametric definition removes the need for assuming a surface brightness profile. Measuring the half-light radius based on SExtractor and/or Curve of Growth methods is also widely used \citep[e.g.][]{ferguson2004,bouwens2004,hathi2008b,oesch2010,curtis-lake2014}. However, these non-parametric methods require a center and an aperture definition to be computed. Such quantities can, again, be misleading in cases of highly disturbed morphology. 

With this in mind we developed a simple size estimator based on the total area covered by a galaxy above a given surface brightness threshold.
In the following sections we define a non-parametric size measurement independent of any assumption related to a symmetric light distribution. Despite being more affected by noise when going towards low surface brightness, this measurement has the advantage of being independent of the shape of the galaxy, and nicely complements the standard $r_e$ measurements.  Other advantages of this method is it does not depend on initial guesses and does not require a fit to converge.

  \subsection{Method}\label{sec:rtot_method}

We define the area of a galaxy by counting the number of pixels above a given surface brightness threshold that belong to it. This approach has the advantage of being completely independent of any asymmetries that the galaxy may have and does not require any centroid/aperture definition.  As most of the objects in our sample show irregular morphology we define physical sizes using the total area $T_x$ as \citep[adapted from][]{law2007}
\begin{equation}
  T_x = N_x L^2 \left( 2\times10^{-11} \frac{\mathrm{ster}}{\mathrm{arcsec^{2}}}\right) D_{A}^{2},
  \label{eq:npsize}
\end{equation}
where $N_x$ is the number of pixels of size $L$ (in arcsec/pixel) that sum up to $x\%$ of the galaxy measured flux and $D_A$ is the angular diameter distance derived using cosmological parameters. From this one can define the area equivalent radius, $r_T^{x}$, 
\begin{equation}
r_{T}^{x} = \sqrt{\frac{T_{x}}{\pi} }.
\end{equation}

In the following we use $T_{100}$ and $r_{T}^{100}$, the area and the equivalent circularized radius that enclose 100\% of the measured flux above a given surface brightness threshold. The quantity $r_{T}^{50}$,  the radius equivalent to the area that sums up to 50\% of the total galaxy light, is similar to the effective radius derived from the GALFIT profile fitting, but taking into account irregular and asymmetric morphology. We also use the values of $r_{T}^{80}$ and $r_{T}^{20}$ enclosing 80\% and 20\% of the flux, respectively.

In practice, the method works as follows. First, we create a segmentation map of a pre-smoothed (Gaussian kernel with 1 pixel width) image stamp of $6\arcsec \times6\arcsec$ centered on the target coordinates. The original image is smoothed so that one can more accurately connect pixels in the faintest region of a galaxy which would otherwise be left out. From the segmentation map we select the area of connected pixels which contains the brightest pixel within a 0.5 arc-second radius from the target coordinates. Using the pixels selected, we compute the total flux of the galaxy by simply summing up the flux in each pixel. Finally, we iterate (using 1000 steps) $f_i$ from the maximum to the minimum flux detected within the segmentation map. At each step we compute the summed flux from the pixels for which $f_\mathrm{pixel}>f_i$. When the summed flux corresponds to $x\%$ of the galaxy total flux, we store the number of pixels that have fluxes greater than $f_i$ at that step and then compute $T_x$ using equation \ref{eq:npsize}.

The selection method chosen for the purpose of this project has the main goal of minimizing the effect of neighbor contamination and of not imposing any size constraints on the derived measurements. We have tested two other different methods: selecting the largest area or the brightest area (by summing the fluxes of all its pixels) for which a subset of its pixels is inside the 0.5'' radius. Both methods yield similar results.

We have tested the  contamination by neighbors dropping simulated galaxies in random regions of the images with and without a constraint on the presence of sources within 2.0 arcsec of  the simulated galaxy. When no constraints are imposed sizes are sometimes artificially increased by the area occupied by a neighboring source. Still, the rate of contamination does not dominate the measurements. We find that the percentage of simulated galaxies with recovered sizes twice as big as the input is of 8\% for the random clean regions and 17\% for the purely random regions. We expect that the contamination rate of our galaxies will be closer to the first reported value since, due to the data processing for a spectroscopic survey, galaxies with large bright companions are excluded from the sample due to problematic redshift estimation. 


\begin{figure*}
\centering
\includegraphics[width=\linewidth]{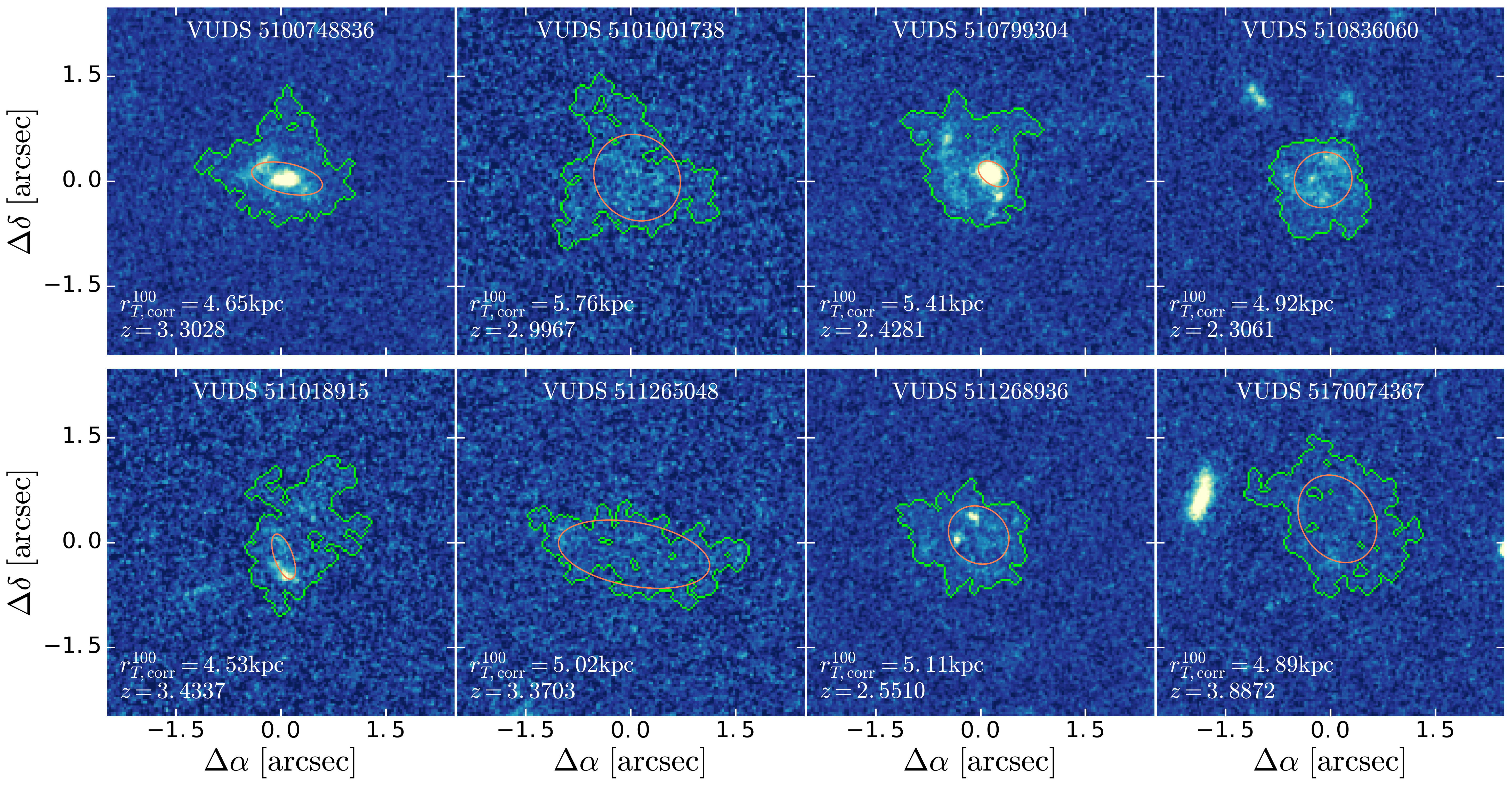}
\caption{F814W stamps from COSMOS \citep{koekemoer2007} of the largest galaxies (all galaxies with $r_{T}^{100,\mathrm{corr}}=4.5$kpc) in our sample at all redshifts. The green contours and red ellipses have the same meaning as in figure \ref{fig:clumpy_galaxies}}
\label{fig:galaxy_example}
\end{figure*}

      \subsection{Cosmological dimming and Luminosity Evolution} \label{sec:rtot_thresh}

\citet{tolman1930} demonstrated that in an expanding universe the relation between the flux and angular size of a nebula should follow the relation (adapted from his equation 30)
\begin{equation}
\frac{F}{\delta \theta^2} \propto \left(1+z\right)^{-4}.
\label{eq:tolman}
\end{equation}
At the time \citep[see][]{lubin2001} it was the only observational test that was proposed to confirm the reality of the expansion of the universe, which is now well established. This dimming effect, independent of the cosmological model, has to be taken into account if one wants to measure galaxy sizes (by means of counting pixels) at the same surface brightness level at different redshifts.

However, when one is analyzing the image recorded with the detector, one is looking at the number of detected photons arriving from the observed source, and not at an integrated flux. Let us consider a square region of the emitting source corresponding to the pixel size at the location of the object, the monochromatic energy emitted is given by
\begin{equation}
L_p = \int_A I(x,y) \mathrm{d}x \mathrm{d}y
\end{equation}
which, for a source emitting at constant brightness inside that region simplifies as
\begin{equation}
L_p = T\times I,
\end{equation}
where $T$ is the physical size of that region from a pixel with a scale $p$:
\begin{equation}
T \propto (p\times D_A)^2  \left(\frac{\mathrm{ster}}{\mathrm{arcsec^{2}}}\right).
\end{equation}
At a distance $D_L$ the observed light is 
\begin{equation}
F_p=\frac{L_p}{4\pi D_L^2}.
\end{equation}
The number of observed photons is
\begin{equation}
N_{\mathrm{photons},o} = \frac{F}{E_{\mathrm{photon},o}},~~E_{\mathrm{photon},o}=h\nu_o=\frac{h\nu_e}{1+z}.
\end{equation}
Combining the equations above we get
\begin{equation}
N_{\mathrm{photons},o} \propto \frac{ I \times (p\times D_A)^2}{4\pi D_L^2}  \frac{(1+z)}{h\nu_e}.
\end{equation}
With $D_L = D_A (1+z)^2$, the number of observed photons per unit area is
\begin{equation}
N_{\mathrm{photons},o} \propto  \frac{p^2}{4\pi} \frac{I}{h\nu_e} \frac{1+z}{(1+z)^4} \propto (1+z)^{-3}.
\end{equation}
The factor $I/(h\nu_e)$ represents the number of emitted photons per unit area. The same redshift dependence was found by \citet{giavalisco1996} and used in \citet{law2007}.

On top of the dimming effect, the average evolution in luminosity of a population of galaxies has an impact on the typical threshold defined at any given redshift. This means that a galaxy with a fixed size but with an increased luminosity because of luminosity evolution will appear larger above a fixed luminosity threshold. 
{The average luminosity evolution from $z\sim4.5$ to $z\sim2$ as reported in the literature is 0.4 magnitudes in the UV rest-frame \cite{reddy2009,bouwens2015}.
}
We take this effect into account parametrizing the evolution of the characteristic luminosity $L_*$ within the redshift range we are considering. Taking the $L_*$ values in \cite{reddy2009} at $z<3$ and those from \cite{bouwens2015} at $z>4$ we parametrize a broken luminosity evolution which is steeper below $z<3$ and flattens at $z>4$:
\begin{equation}
L(z)=
\begin{cases} 
       10^{-0.4(-0.36z)} & z\leq 3 \\
      2.25\times 10^{-0.4(-0.07z)} &z>3 \\
   \end{cases}
\label{eq:lumevo}
 \end{equation}

 \subsection{Detection threshold}
 
The dimming effect described in the previous subsection relates to an important tuning parameter for this non-parametric size measurement which is the detection threshold above which a pixel is classified as a detection and may be assigned to the galaxy. This detection threshold is defined as a multiple $k$ of the standard deviation of the sky emission, $\sigma$, computed globally. To be sure that we are fairly comparing galaxies imaged at different depths, the value of $\sigma$ is taken as the median standard deviation of the image noise computed from 100 different regions in the shallowest survey. In our case, $\sigma=1.1\times10^{-14}\ \mathrm{ergs\ s^{-1}cm^{-2} arcsec^{-2}}$ 
is computed from the COSMOS images. 

The choice of $k$ plays an important role on the size estimates that one gets. The simplest assumption is to define a constant value $k_0$ across all redshifts. Another approach consists in taking into account the effect of cosmological dimming and typical luminosity evolution (described above) and choose $k$ as a function of the galaxy redshift following
\begin{equation}
k = k_p \left( \frac{1+z}{1+z_p} \right)^{-3} \times \frac{L(z)}{L(z_p)},
\label{eq:thresh}
\end{equation}
where $k_p$ is the value of $k$ at the pivot redshift $z_p$ and $L(z)$ is defined in equation \ref{eq:lumevo}. To use this method, the availability of spectroscopic redshifts is even more important as errors in the redshift would affect the threshold used and consequently impact the final size measurements of the galaxies.

          \begin{figure}
   \centering
   \includegraphics[width=\linewidth]{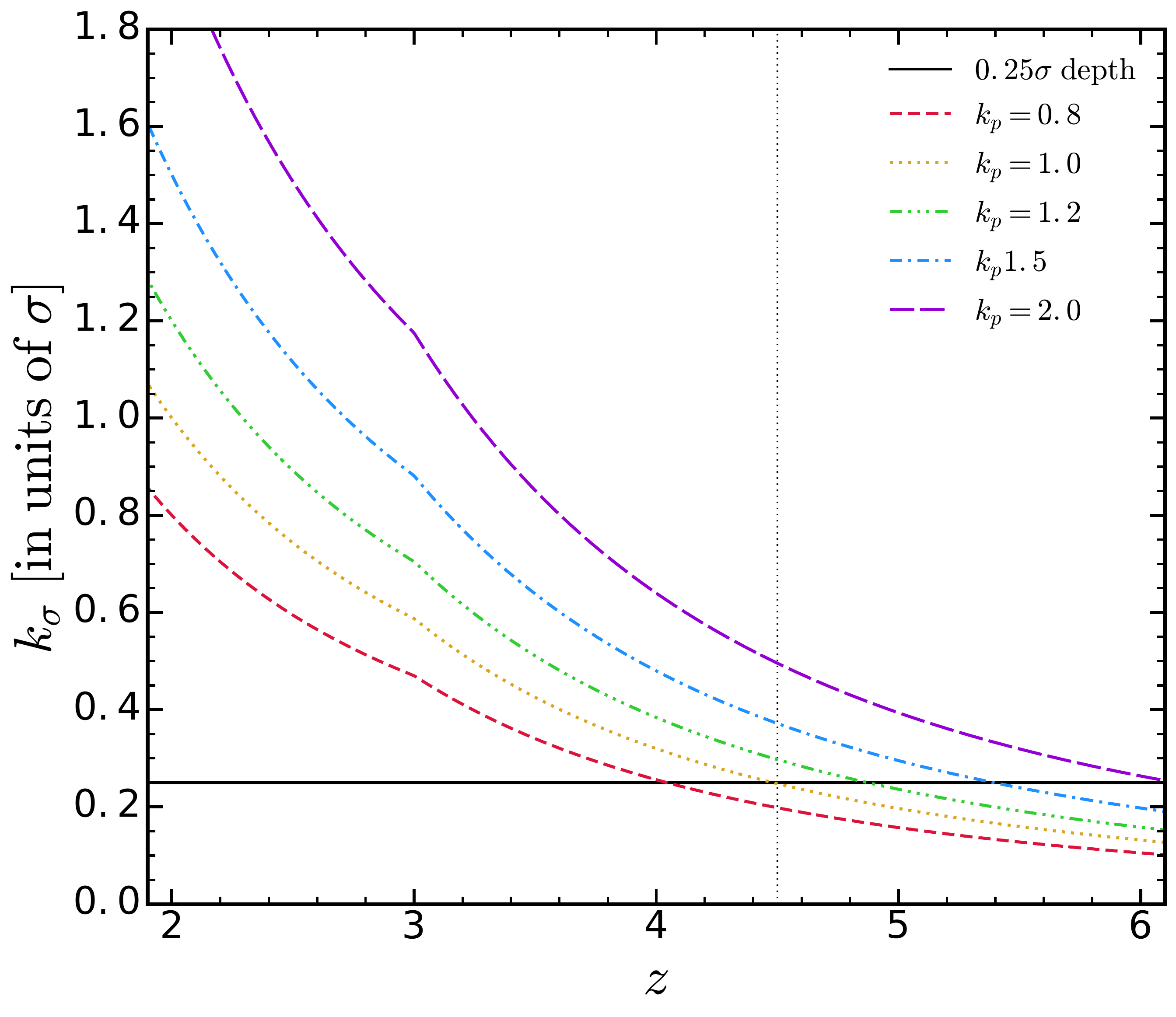}
      \caption{Value of $k$ as a function of redshift for different choices of $k_p$. The vertical dotted line shoes the upper redshift limit we assume in this paper. The black solid line shows the depth limit above which the noise correction is larger than 100 pixels.}
         \label{fig:kthreshold}
   \end{figure}

Such drastic evolution of the detection threshold (see figure \ref{fig:kthreshold}) led us to restrict the redshift interval of our study to $2<z<4.5$. In this range we probe the total extent of galaxies down to a lower surface brightness level at lower redshift while avoiding to overestimate galaxy sizes at the high redshift end due to random sky background noise detection. To obtain a size measurement corresponding as closely as possible to the total physical area covered by a galaxy, the choice of the threshold is based on the average number of connected sky pixels as a function of the threshold. We find that it is at the level of $1.0\sigma$ that we start to detect connected pixels (4-6 pixels) only due to random sky noise fluctuations. Based on that, and on the average galaxy color which we discuss in section \ref{sec:rtot_color}, we define $k$ such that at $z=2$ it has a value of $1.0$ to probe galaxies down to their fainter regions. This corresponds to a detection limit of $\approx 0.25\sigma$ per pixel at $z=4.5$. We discuss the contamination from random fluctuations and the correction made to the size measurements measured down to these thresholds in section \ref{sec:rtot_sky}.

The galaxy sizes that we measure are tied to the choice of $k_p$ and/or the depth of the image which determines the value of $\sigma$. By setting the value of $k \times \sigma$ at higher/lower values, the absolute value for the galaxy size will decrease/increase. We therefore report total size measurements indexed to a specific choice of $k_p$, i.e. to a specific limiting isophote defined at $z=z_p$. For the sake of simplicity we shall refer hereafter to our non-parametric size as the total size.

%


  \subsection{Correcting for PSF broadening}\label{sec:rtot_psfs}
Since we are measuring  sizes directly on  images we are affected by the broadening of the profile by the instrumental PSF. To correct for this effect we simulate with GALFIT a PSF image re-scaled to match the magnitude of the galaxy. We then run the simulated PSF through the same detection algorithm and obtain the size of the PSF at the object magnitude and same surface brightness level. It is then straightforward to compute the galaxy size as
\begin{equation}
r_{T,\mathrm{psf-corr}}^{x} = \sqrt{ \left(r_{T}^{x}\right)^2 - \left(r_{T,\mathrm{PSF}}^{x}\right)^2 } = \sqrt{\frac{T_x - T_{x,\mathrm{PSF}}}{\pi}}
\label{eq:rT_definition}
\end{equation}


      \subsection{Dependency with color}\label{sec:rtot_color}

The method of measuring galaxy sizes presented above is dependent of individual pixel detection, and it is therefore more sensitive to the galaxy spectral energy distribution  than other methods relying on a given surface brightness profile. Even in the absence of color gradients, the simple fact that one object could  have a wavelength-dependent brightness leads the number of selected pixels to vary due to relative brightness change of the galaxy within a given filter with respect to the detection threshold.

To correct for this we apply a simple color correction to the detection threshold taking into account the different brightness at different rest-frame wavelengths: 
\begin{equation}
k_c = k\times 10^{-0.4(I_\mathrm{obs}-NUV_\mathrm{rest})},
\label{eq:thresh_color}
\end{equation}
where $k$ is that of equation \ref{eq:thresh} and $(I_\mathrm{obs}-NUV_\mathrm{rest})$ is the color term. The color is defined to be the observed band minus the rest-frame $NUV$ band derived from SED fitting with Le Phare. The choice of the $NUV$ band was done such that, for the majority of our galaxies, the color term is minimal since it overlaps with the observed band.



  \subsection{Correcting for sky pixels detection}\label{sec:rtot_sky}
  
In the presence of uncorrelated Gaussian noise, the probability of a \emph{sky} pixel being above the detection threshold defined above is not negligible. To estimate the level of contamination produced by random sky noise, we compute the $N_{100}$ area (equation \ref{eq:npsize}, size of the segmentation map) on a search radius of  0.5 arc second (the same as when computing $T_{100}$ for the real galaxies) randomly selected on 500  $300\arcsec\times300\arcsec$ regions. 

For each search radius, we compute the area defined for surface brightness thresholds ranging from 0.05$\sigma$ to $2.05\sigma$ using steps of 0.025. This provides the median area of sky connected pixels for a given threshold value. We use this to correct  the value of $r^{100}_T$ following 
\begin{equation}
r_{T,\mathrm{corr}}^{100} = \sqrt{ \left(r_{T,\mathrm{psf-corr}}^{100}\right)^2 - T_\mathrm{sky}(k)/\pi }
\label{eq:rTskycorr_definition}
\end{equation}
where $T_\mathrm{sky}(k)$ is the median number of connected \emph{sky} pixels at a threshold $k$. In figure \ref{fig:simulated_T100} we show the impact of this correction on a set of simulated galaxies. We stress that this correction is only applied to the value of $r^{100}_T$. We note however that the impact of the noise is greater when increasing the fraction of the flux used to derive sizes. Sky noise effects are negligible for sizes computed using $10\%$ of the total flux.

  \subsection{Results}\label{sec:rtot_results}


Of the 1242 galaxies in our stellar mass selected sample, 67 objects have no measured value of $r_T$. These failures are due to the lack of pixel fluxes above the selection threshold defined in equation \ref{eq:thresh_color} for 17 objects and due to the lack of color information for 5 objects. The other 45 objects are on the edge of the image mosaic and thus no sufficient information is available to derive a size. 

\begin{figure}
   \centering
   \includegraphics[width=0.49\textwidth]{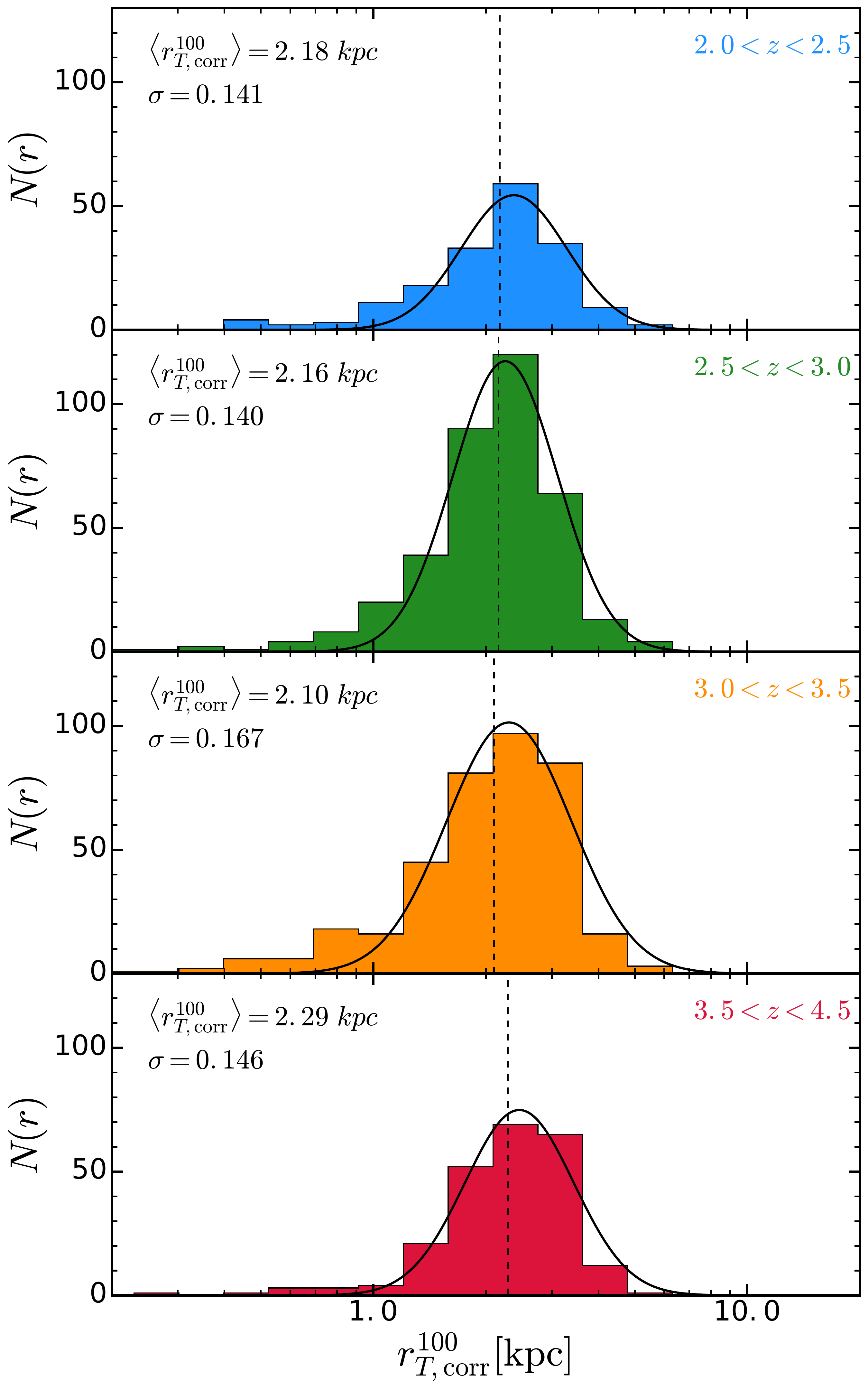}
      \caption{Total size distribution for different redshift bins. The vertical dotted line indicates the median value for each bin. The black line is the fit of the distribution using a normal function in log space.}
         \label{fig:rT100_distribution}
   \end{figure}

The distribution of total sizes is presented in Figure \ref{fig:rT100_distribution} for the four redshift bins previously defined. The shape of the distribution remains roughly the same with redshift. The Gaussian fit on each sub-sample yields sigma values of (0.141,140,0.167,0.146) with increasing redshift.

\begin{figure}
   \centering
   \includegraphics[width=\linewidth]{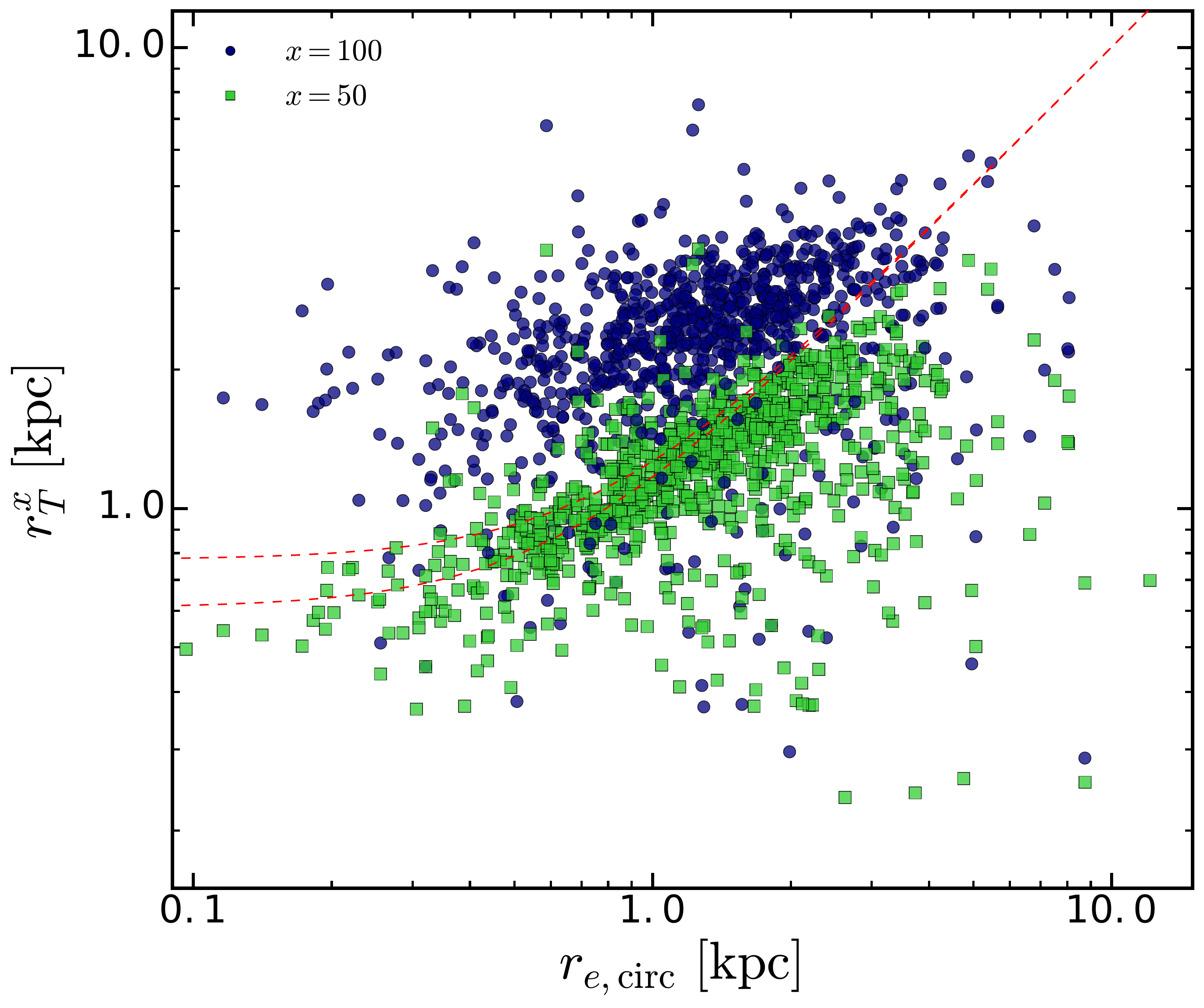}
      \caption{Comparison of the two size estimates computed in this paper. In dark blue points we have the comparison to $r_{T,\mathrm{corr}}^{100}$ and in green points to $r_T^{50}$. The red dashed lines corresponds to the equation $r_T = \sqrt{r_{e,\mathrm{circ}}^2+\mathrm{FWHM}^2}$ for the limit redshifts 2 and 4.5 (top and bottom lines respectively).}
         \label{fig:rT_re}
   \end{figure}

We compare the total size measurements to the circularized effective radius obtained in Section \ref{sec:re_results} in Figure \ref{fig:rT_re}. Galaxies with the larger $r_{e,\mathrm{circ}}$ are also larger in total size. However, this trend deviates from the linear relation and there is a large scatter in the relation. The scatter is somewhat expected as we are dealing with highly irregular morphology. While GALFIT gives a heavier weight to the brightest pixels (which can be locked into a single bright clump in a large galaxy), $r_T$ measures the whole galaxy extent. 
GALFIT can pick up fainter \emph{buried} flux of a galaxy by extrapolation of the profile measured on the bright flux, which is not picked up when computing the total size using our method which explains why some galaxies have $r_{e,\mathrm{circ}}$ larger than $r_T$.
It is interesting to note that the $r^{50}_{T}$ is nicely correlated with $r_{e,\mathrm{circ}}$, as shown in Figure \ref{fig:rT_re}, with a dispersion of 0.25 dex which is a consequence of the frequent asymmetric shapes in our sample. 

\begin{figure}
   \centering
   \includegraphics[width=\linewidth]{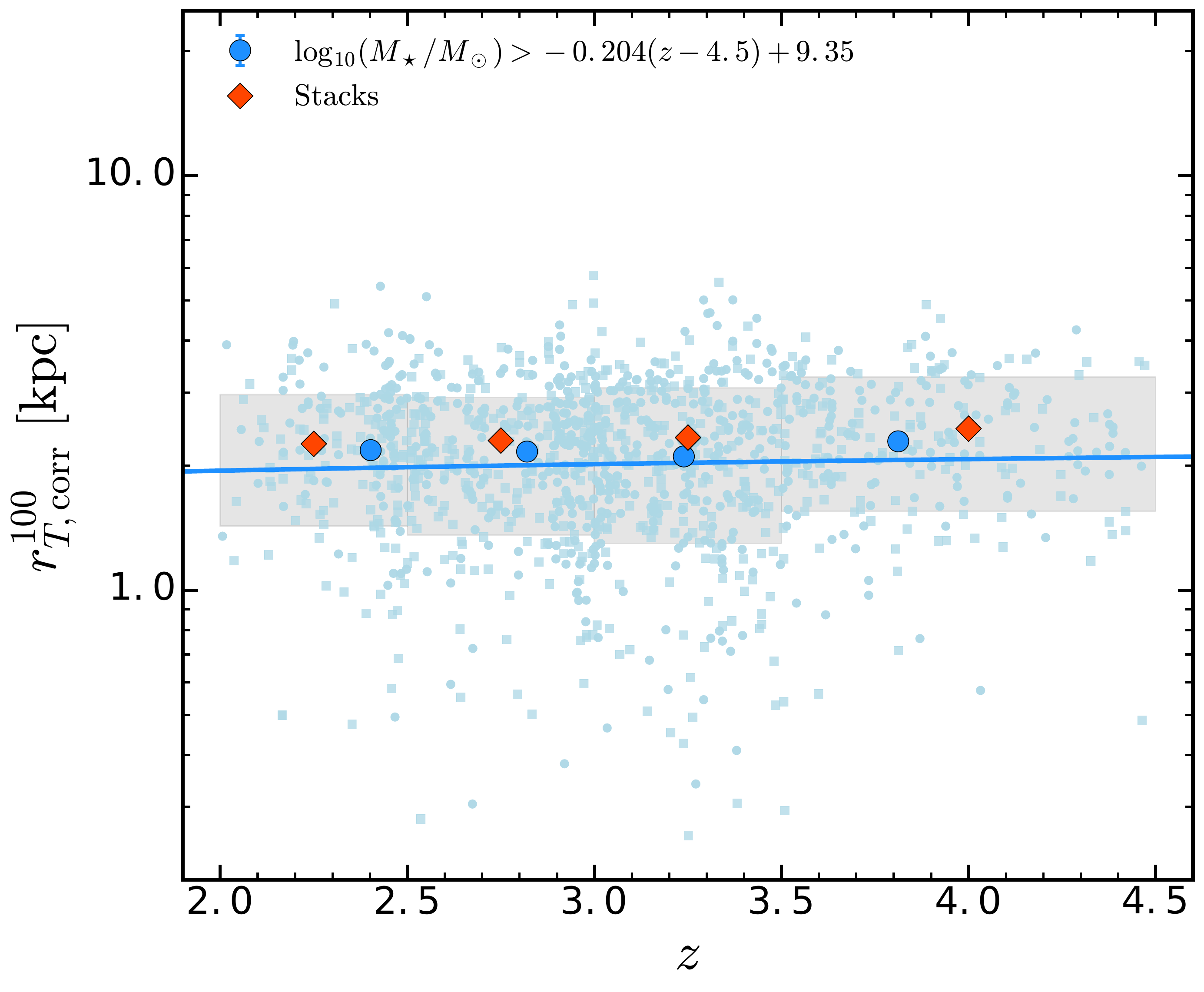}
      \caption{Total size evolution with redshift. Each galaxy is plotted with a small blue point (squares for redshift confidence 2 and 9, circles for 3 and 4). The median values and respective error ($\sigma/\sqrt{N}$) per redshift bin are shown by the large blue points with the error bars. The shaded region delimits the 16th and 84th percentiles including 68\% of the sample in each redshift bin. The red diamonds are from the stacked images computed from the method described in section \ref{sec:stacks}.}
         \label{fig:rT100_evolution}
   \end{figure}

The evolution of total sizes across the redshifts considered here is shown in figure \ref{fig:rT100_evolution}.
We find that total sizes are staying roughly constant over the redshift range $2<z<4.5$, with $r^{100}_{T} \simeq 2.2$ kpc.  The dispersion around the mean of $\sigma \simeq 0.9$ kpc (0.21 dex), and we find galaxies that are as large as $\sim5.5$ kpc and as small as $\sim0.5$ kpc at all redshifts in this range. Size measurements using this definition are consistently larger than sizes provided by the effective radius $r_e$. This is discussed in Sections \ref{sec:size_evolution} and \ref{sec:discussion}.

\begin{figure}
   \centering
   \includegraphics[width=\linewidth]{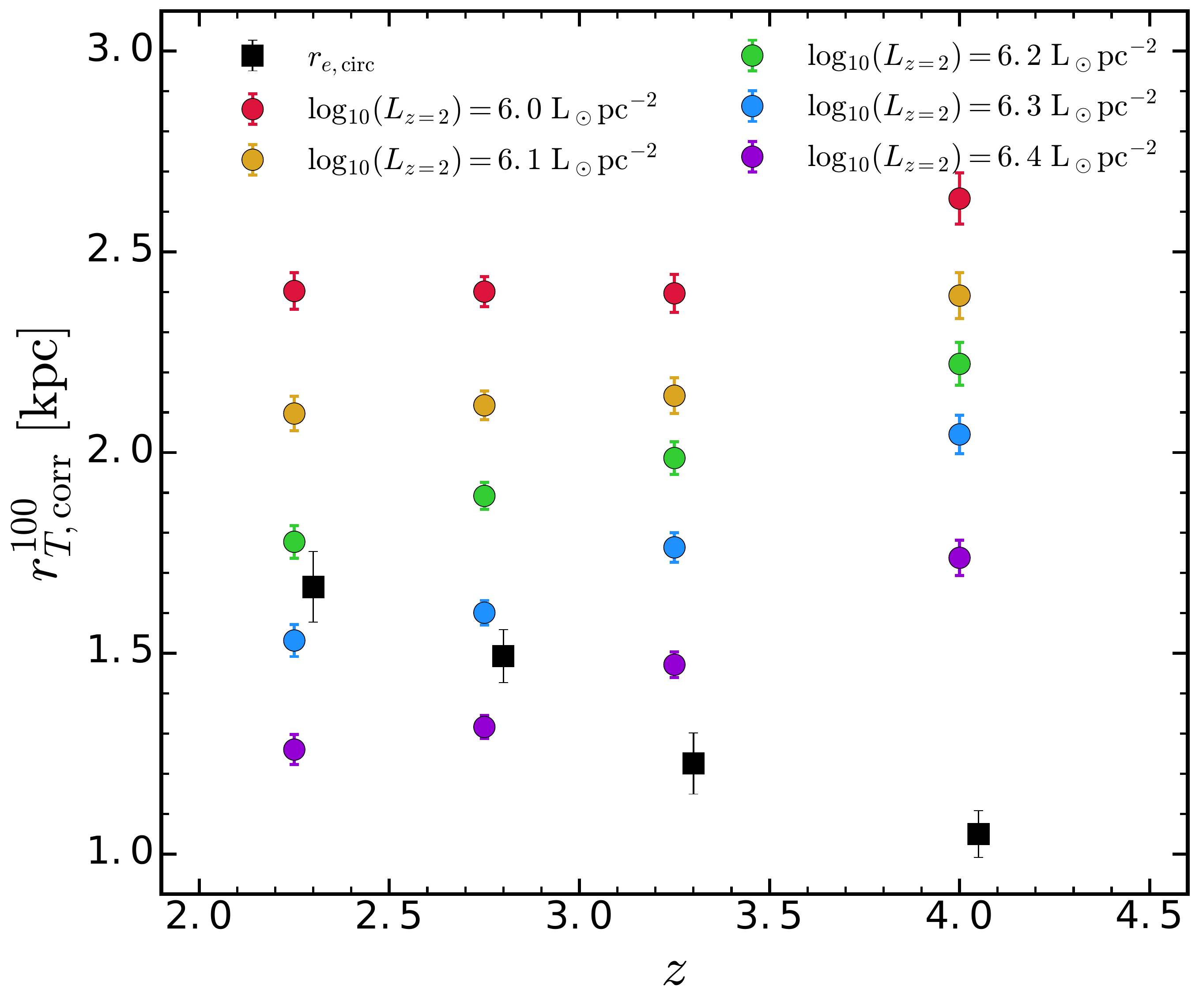}
      \caption{Total size measurements defined at different values of $k_p$ that correspond to different luminosity thresholds at $z=2$. From top to bottom, the colored circles correspond to $k_p=0.8,1.0,1.2,1.5,2.0$. The color coding is the same as figure \ref{fig:kthreshold}. }
         \label{fig:rT100_evolution_multithresh}
   \end{figure}

We show the dependence of the absolute value of the measured sizes as a function of the chosen value of $k_p$ in figure \ref{fig:rT100_evolution_multithresh}. As expected when using different limiting isophotes over a range of 2.5 in luminosity changes the size values. It is  therefore essential to quote the absolute luminosity of the isophote used to compute galaxy sizes. We note that at all times, the observed trends as a function of redshift are consistently different from those observed with GALFIT, with approximately constant sizes as a function of redshift while the effective radius is observed to strongly decrease.

\section{Average galaxy profiles and light concentration}\label{sec:stacks}

Since the total sizes of galaxies, measured at the same surface brightness level, seem to not evolve much with redshift, we investigate if the same pattern applies to the light profiles. We do this by looking at the stacked profiles and at the individual measurements of light concentration in galaxies.

      \subsection{Image stacks}\label{sec:stacks_method}

In order to investigate the evolution of the light profiles  we stack images of the large number of sources in our sample to follow the profile at larger radii than from individual images. The stacking procedure starts by collecting a $10\arcsec\times10\arcsec$ stamp image for each galaxy in the sample. Then, one runs SExtractor to find the light-weighted center of each source. We shift the image by the difference between the light-weighted center and the physical image center, using a cubic spline interpolation to re-scale the image onto a common grid. After processing the entire sample for a given redshift bin, we take, at each position, the median pixel value as the \emph{flux} of the stacked image. We also produce a straightforward sum to the images normalized by the total galaxy flux. The results of both methods give the same qualitative answers.

Stacking within a redshift bin, smears the rest-frame light over the interval. Also, when comparing different redshift bins we are effectively looking at different rest-frame light. This ranges from $\sim2500\AA$ in the lowest redshift bin down to $\sim 1600\AA$ in the highest redshift bin. As shown in section \ref{sec:re_color}, sizes derived  are roughly the same across this wavelength range. This difference in rest-frame wavelengths has no effect on the comparison of the profiles discuseed in the next section.


           \begin{figure}
   \centering
   \includegraphics[width=\linewidth]{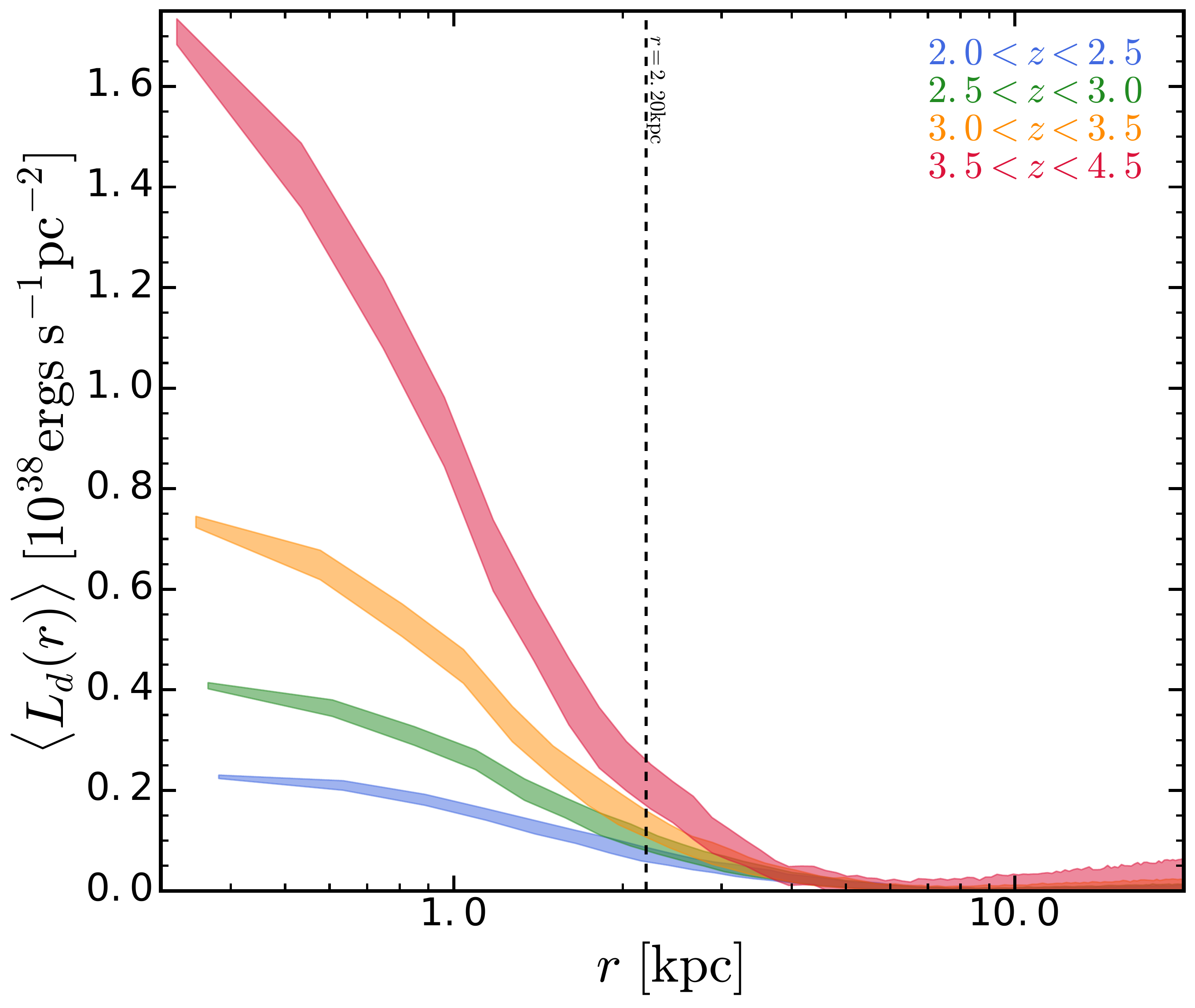}
      \caption{Surface luminosity profiles of the median stacked F814W band images of galaxies in our sample at four different redshift bins. The black dashed line is the median $r_{T,\mathrm{corr}}^{100}$ of our sample. The width of each profile encodes the surface luminosity error.}
         \label{fig:median_stack_sbp}
   \end{figure}

      \subsection{Luminosity profiles}\label{sec:stacks_evolution}

The redshift dependency of the light profile derived from the stacked images is presented in Figure \ref{fig:median_stack_sbp}. We convert the median stacked profiles into surface luminosity profiles and show these profiles without any normalization. The light profiles change smoothly across redshift starting from a highly peaked, concentrated profile at the higher redshifts $z \sim 4$-$4.5$ and becoming more smoothed and less concentrated as we move towards the lower redshifts $z\sim2$-$2.5$.

The strong evolution observed here does not seem to come from a limitation from the observations, as noise contribution would tend to produce the opposite trend to what is observed, with flatter profiles at high redshifts rather than the more peaked profiles that we observe. Indeed, if the light profiles were similar in shape and measurements on faint galaxies were being increasingly noise dominated at the highest redshifts, the light profiles at high redshifts would be flatter than at low redshifts: the sky residuals would tend to flatten the outer wings of the profiles effectively reducing the steepness of the overall profile. 
      
      \subsection{Light concentration of individual galaxies}\label{sec:stacks_concentration}

To further investigate the change in light profiles with redshift we use a different but quite commonly used way to quantify the light concentration. The concentration parameter was originally defined by \citet{bershady2000} as the ratio of the 80\% and 20\% light radius derived from elliptical apertures centered on the galaxy. This measurement assumes symmetrical isophotes to derive the galaxy concentration, which makes it difficult to apply to irregular and asymmetric galaxies at high redshifts as it is the case of our sample. To overcome this limitation we derive the concentration parameter $C_T$, using $r_T^{20}$ and $r_T^{80}$ as defined in Section \ref{sec:rtot}, generalizing the computation of $C$ to the case of irregular and asymmetric galaxies
\begin{equation}
C_T = 5\times \log_{10}\left( \frac{r_{80}^{T}}{r_{20}^{T}} \right).
\label{eq:concentration}
\end{equation}

\begin{figure}
\centering
\includegraphics[width=\linewidth]{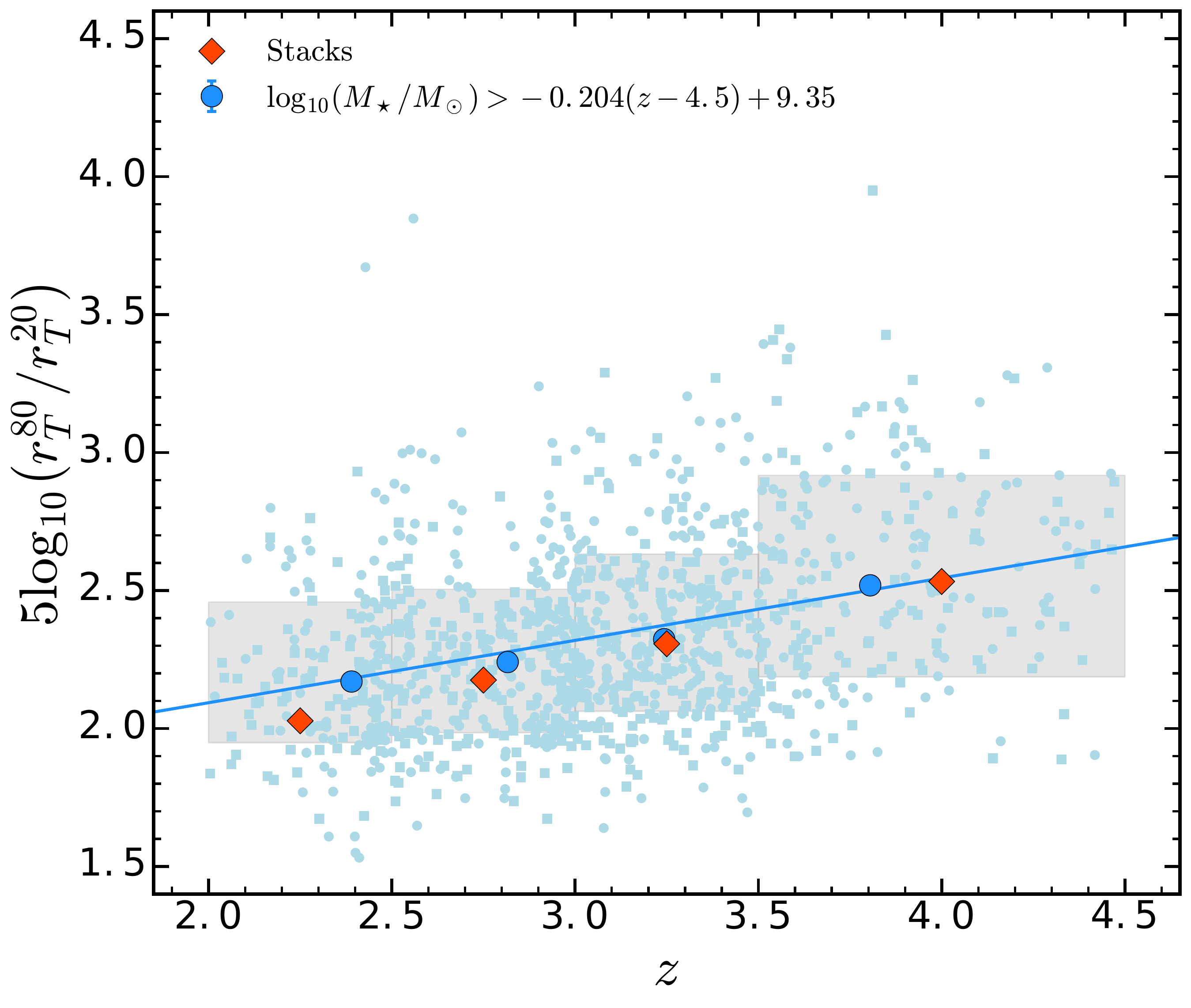}
\caption{Evolution of light concentration in galaxies computed from the ratio of the non-parametric radii enclosing 80\% and 20\% of the total flux, as defined in Section \ref{sec:rtot}. The solid line is a linear fit to all the small blue points.Each galaxy is plotted with a small blue point (squares for redshift confidence 2 and 9, circles for 3 and 4). The median values and respective error ($\sigma/\sqrt{N}$) per redshift bin are shown as  large blue points with error bars. The shaded region delimits the 16th and 84th percentiles including 68\% of the sample in each redshift bin. The orange diamonds are the values obtained from the stacked images computed from the method described in section \ref{sec:stacks}.}
\label{fig:concentration}
\end{figure}

We plot the evolution of $C_T$ in figure \ref{fig:concentration}. In that figure we show the fit of the equation
\begin{equation}
C_T = \alpha_C z+\beta_C
\end{equation}
and find a value of $\alpha_C=0.23 \pm 0.02$ and $\beta_C=1.64\pm0.06$. Thus, $C_T$ strongly evolves with redshift, light being more concentrated going to higher redshifts. The concentration $C_T$ increases by $\sim25$\% from $z\sim2$ to $z\sim4.5$.

Using the standard definition of concentration with symmetric apertures (see appendix \ref{app:concentration} for details on the computation) leads to an different trend ($\alpha_C=0.01\pm0.04$ and $\beta=2.82\pm0.13$) with light concentration being rather constant within our redshift interval. This difference is easily explained because of the impact of defining a galaxy center for irregular morphology. For instance, on a galaxy composed by several same-brightness level clumps, the center of light would be defined somewhere in between the clumps leading to a lower concentration as the radius containing 20\% of the total light would be larger to include the clumps. 

The concentration values derived for the median stack images report the same trend as those derived from the full sample (figure \ref{fig:concentration}). While the concentration value for the stacked images is sensitive to the threshold definition (average color and $k_p$ from equation \ref{eq:thresh}) the evolution trend remains unaffected and the slope is rather stable.

\section{Size evolution with redshift}\label{sec:size_evolution}

On sections \ref{sec:re_results} and \ref{sec:rtot_results} we have mentioned that we could not obtain galaxy sizes for the entirety of our sample for each of the measurements presented and for different reasons in each case. We have tested a scenario where we remove from the sample all galaxies without both size measurements which comprise 238 (19\%) of the stellar mass selected sample. We find that the results do not change due to the removal of those objects and thus we kept them in the final sample for the individual analysis of $r_T$ and $r_e$ to reduce the statistical errors in our measurements.

\begin{table*}
\centering
\begin{tabular}{|c|c|c|c|c|}
\cline{2-5}
\multicolumn{1}{c|}{} &\multicolumn{1}{|c|}{$2.0<z<2.5$}&\multicolumn{1}{|c|}{$2.5<z<3.0$}&\multicolumn{1}{|c|}{$3.0<z<3.5$}&\multicolumn{1}{|c|}{$3.5<z<4.5$}\\
\hline
$r_e$ &1.67 $\pm$ 0.09&1.49 $\pm$ 0.07&1.23 $\pm$ 0.08&1.05 $\pm$ 0.06\\
$r_T^{50}$ &1.30 $\pm$ 0.04&1.25 $\pm$ 0.02&1.13 $\pm$ 0.02&1.25 $\pm$ 0.03\\
$r_{T,\mathrm{corr}}^{100}$ &2.18 $\pm$ 0.06&2.16 $\pm$ 0.04&2.10 $\pm$ 0.05&2.29 $\pm$ 0.05\\
\hline
\end{tabular}
\caption{Median size values (for F814W, in kiloparsecs) per redshift bin considered here. The value of $r_e$ is the circularized effective radius from GALFIT, $r_T^{50}$ corresponds to the PSF corrected values and $r_{T,\mathrm{corr}}^{100}$ to the psf+sky corrected values.}
\label{tab:sizes}
\end{table*}

\begin{table*}
\centering
\begin{tabular}{|c|cc|cc|}
\cline{2-5}
\multicolumn{1}{c|}{} &\multicolumn{2}{|c|}{F814W}&\multicolumn{2}{|c|}{F160W}\\
\hline
Size estimate & $\alpha_r$ & $\beta_r$ & $\alpha_r$ & $\beta_r$ \\
\hline
$r_e$ & -1.29  $\pm$ 0.18  &  0.89 $\pm$  0.11 & -1.40 $\pm$  0.34 &  1.01 $\pm$  0.20 \\
$r_T^{10}$& -0.44 $\pm$ 0.09 & -0.12 $\pm$ 0.06 & -0.44 $\pm$ 0.09 & -0.12 $\pm$ 0.06 \\
$r_T^{20}$ &-0.40 $\pm$ 0.10 & 0.02 $\pm$ 0.06 & -0.18 $\pm$ 0.22 & -0.04 $\pm$ 0.13 \\
$r_T^{30}$ &-0.36 $\pm$ 0.10 & 0.12 $\pm$ 0.06 & -0.09 $\pm$ 0.23 & -0.01 $\pm$ 0.14 \\
$r_T^{40}$ &-0.29 $\pm$ 0.11 & 0.17 $\pm$ 0.06 & 0.03 $\pm$ 0.26 & -0.01 $\pm$ 0.16 \\
$r_T^{50}$ &-0.21 $\pm$ 0.10 & 0.19 $\pm$ 0.06 & 0.13 $\pm$ 0.20 & 0.01 $\pm$ 0.13 \\
$r_T^{60}$ &-0.11 $\pm$ 0.10 & 0.20 $\pm$ 0.06 & 0.24 $\pm$ 0.22 & -0.02 $\pm$ 0.14 \\
$r_T^{70}$ &-0.01 $\pm$ 0.10 & 0.20 $\pm$ 0.06 & 0.41 $\pm$ 0.25 & -0.08 $\pm$ 0.15 \\
$r_T^{80}$ &0.11 $\pm$ 0.10 & 0.18 $\pm$ 0.06 & 0.56 $\pm$ 0.27 & -0.14 $\pm$ 0.17 \\
$r_T^{90}$ &0.25 $\pm$ 0.10 & 0.15 $\pm$ 0.06 & 0.67 $\pm$ 0.28 & -0.19 $\pm$ 0.17 \\
$r_T^{100}$ &0.38 $\pm$ 0.11 & 0.11 $\pm$ 0.07 & 0.34 $\pm$ 0.28 & 0.07 $\pm$ 0.17 \\
\hline
$r_{T,\mathrm{corr}}^{100}$ &0.13 $\pm$ 0.12 & 0.23 $\pm$ 0.07 & 0.02 $\pm$ 0.24 & 0.24 $\pm$ 0.14 \\
\hline
\end{tabular}
\caption{Best fit slope and normalization factor for equation $\log(r)=\alpha_r\log(1+z)+\beta_r$ for the different size estimates used throughout the paper. Values for both F814W and F160W, valid for $2<z<4.5$, are given here.}
\label{tab:slopes}
\end{table*}

\begin{figure}
   \centering
   \includegraphics[width=\linewidth]{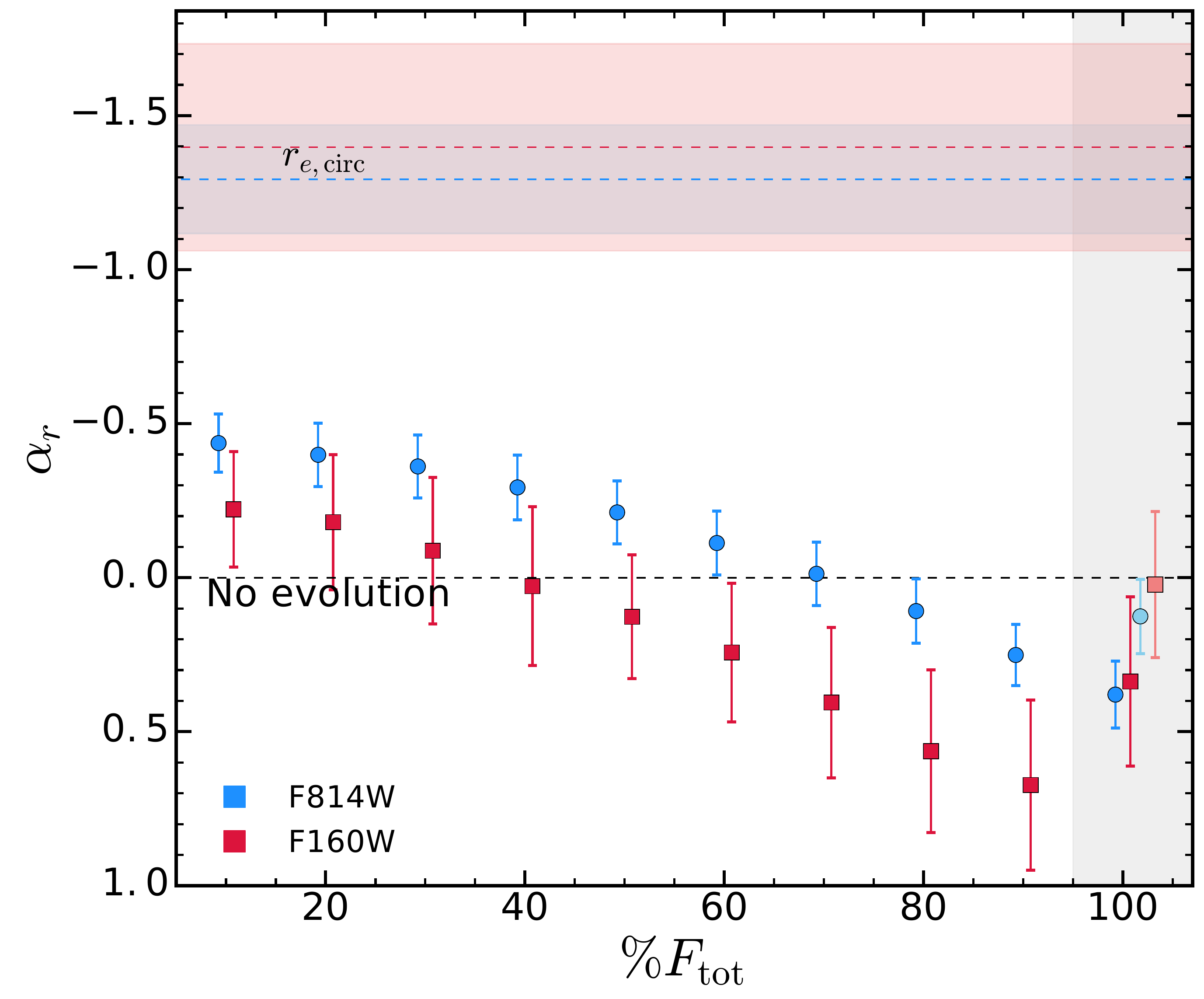}
   \caption{Slope of the evolution of galaxy sizes as a function of the flux level considered. The light colored symbols corresponds to the slope computed from sky-corrected $r_T^{100}$ sizes. The dashed lines and shaded regions are the value and error derived in $\alpha$ for the GALFIT size measurements. Blue and red coloured symbols/regions correspond to F814W and F160W measurements respectively.}
\label{fig:size_evolution_all}
\end{figure}

To quantify the redshift evolution of sizes we fit the data points with a commonly used parameterization \citep[e.g.][]{shibuya2015,straatman2015,vanderwel2014,morishita2014}:
\begin{equation}
\log_{10}(r) = \beta_r+\alpha_r\log_{10}(1+z)
\label{eq:size_evol}
\end{equation}
We fit this relation to all galaxies with a size measurement weighted equally due to the lack of individual errors on $r_T^{x}$. The sample variance of the parameters $\beta_r$ and $\alpha_r$ is obtained from bootstrapping the data 5000 times using 80\% of the full sample each time. The resulting fit through the data are plotted in figures \ref{fig:re_evolution} and \ref{fig:rT100_evolution}. We also compute the same evolutionary trend for different values of $x$ of equation \ref{eq:npsize} ranging form 10\% up to 100\% with a 10\% step value. All the fit results are given in Tables \ref{tab:sizes} and \ref{tab:slopes} and summarized in Figure \ref{fig:size_evolution_all}. 

The Figure \ref{fig:size_evolution_all} encodes the information on the slope of the size evolution as a function of the flux level and compares to the effective radius evolution derived from GALFIT. There are three main points to be taken from this figure. First, there is a clear trend of the slope as a function of the flux cut level: the brightest regions  (10\% brightest pixels) evolve faster than the total size with a smooth trend towards slower evolution when going from the brightest regions to the total area of the objects. This differential trend is again a hint on the evolution of light concentration as a function of redshift where the brightest regions of the galaxy tend to become larger with time while the total size remains constant effectively suggesting that light becomes less concentrated with time. Second, the evolution in sizes is far from that reported from the effective radii $r_{e,\mathrm{circ}}$. For the size measurements including the brightest pixels that amount to 50\% of the total flux (comparable to $r_e$) the derived evolution slopes are separated by more than $3\sigma$ (Figure \ref{fig:size_evolution_all}) indicating that the choice of method to compute sizes can have a very significant impact on the observed evolution. We note that the points in figure \ref{fig:size_evolution_all} indicate the lower limits on the slope. At the highest redshifts a galaxy size will be more often overestimated as lower flux thresholds increase noise contamination. We attempt to correct for that using the prescription described in Section \ref{sec:rtot_sky}, but it applies only to $r_T^{100}$.  However, this effect can not explain the observed difference between $r_e$ and $r_{T}$ as the size evolution slope after the said correction still differs by more that $3\sigma$ from that derived from $r_{e,\mathrm{circ}}$ and it is where the impact of the noise affects the slope the most.

  \subsection{Effective radius $r_e$ evolution}\label{sec:re_evolution}

The galaxy size evolution derived from the circularized effective half-light radius shows strong size evolution across the redshift range probed. Assuming that galaxy sizes evolve according to the parametrization of equation \ref{eq:size_evol} we derive a slope of $\alpha_r=-1.29\pm0.18$.

Values reported in the literature range from $\alpha_r=-0.56\pm0.09$ for a sample of star-forming spectroscopically confirmed galaxies at $0.5<z<3.0$ with $\log_{10}(M_\star/M_\odot)>10$  \citep{morishita2014} up to $\alpha_r=-1.32\pm0.52$ for a sample of LBGs with $(0.12-0.3)L_\star$ at $z\sim4-6$ \citep{oesch2010}. The differences between these results \citep[see also][]{ferguson2004,bouwens2004,trujillo2006a,franx2008,williams2010,mosleh2011,mosleh2012,ono2013,vanderwel2014,morishita2014,curtis-lake2014,shibuya2015,straatman2015} can be traced to different likely sources: 1) the samples are selected in different ways; 2) there are two different methods for measuring sizes, both relying on a certain degree of symmetry; and perhaps most importantly 3) samples with different stellar masses yield different sizes \citep{franx2008,vanderwel2014}.


    \subsection{Total size $r_{tot}$ evolution}\label{sec:rtot_evolution}


The typical total size of a SFG, corrected for sky noise contamination as computed in Section \ref{sec:rtot} ,is $r_T^{100} \simeq 2.2$kpc and has little to negligible evolution of galaxy sizes across the redshifts probed here (see the light color points of figure \ref{fig:size_evolution_all}). This is markedly different from what is observed using the standard effective radii measurement. The $r_T^{100}$ measurements are consistent with the results of \citet{law2007} who report only a small increase in the the total area from $z\sim2$ to $z\sim3$ ($T =15.0\pm0.7$ and $17.4\pm 0.9$ kpc$^{2}$, respectively). This corresponds to an equivalent radius of $r_T = 2.19 \pm 0.47$kpc at $z\sim2$ and $r_T = 2.35\pm0.54$kpc at $z\sim3$.


\section{Size relations with physical parameters}\label{sec:size_physic}

To further explore the properties of galaxies with different sizes we compare galaxy sizes to several key physical parameters characterizing galaxies including stellar mass, SFR, age, metallicity and dust extinction. We also investigate any possible relation between sizes and inter-galactic medium (IGM) transmission towards the galaxies.
These parameters are derived from the simultaneous SED fitting of the VUDS spectra and all multi-wavelength photometry available for each galaxy, using the code GOSSIP+ as described in Thomas et al. (submitted). This method expands the now classical SED fitting technique to the use of UV rest-frame spectra in addition to photometry, further improving the accuracy of key physical parameter measurements (see Thomas et al. for details).  

\begin{figure*}
\centering
\includegraphics[height=0.90\textheight]{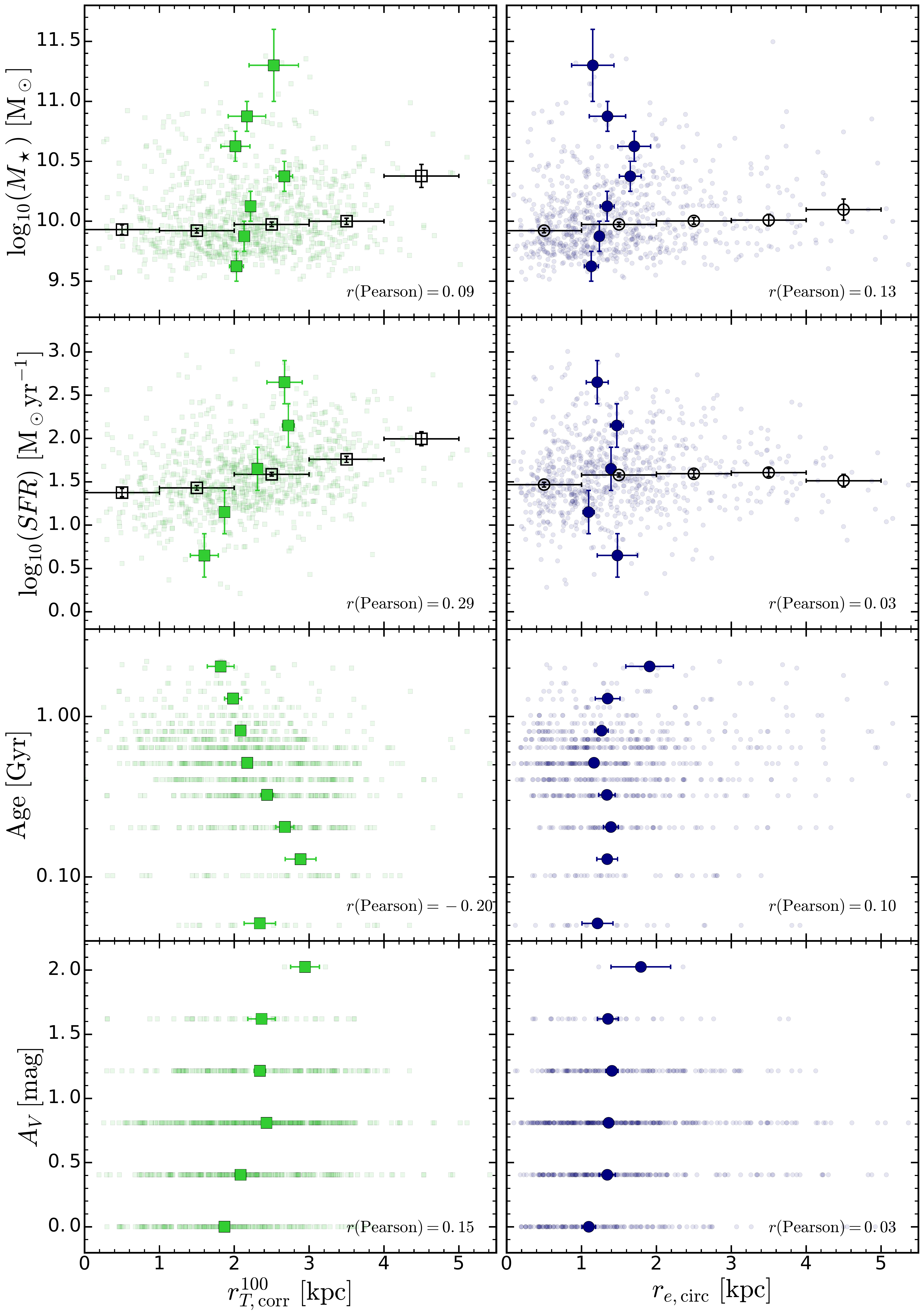}
\caption{Correlation of the measured galaxy sizes with parameters derived from GOSSIP+ SED fitting for our stellar mass selected sample (using the stellar masses derived with GOSSIP+ and equation \ref{eq:mass_sel}). Only galaxies with good and excellent spectrophotometric fits are included. From top to bottom we have the stellar mass, SFR, age and dust extinction. On the right panels we show the results for GALFIT derived $r_{e,\mathrm{circ}}$ and in the left panels we show the results for our value of  $r_{T,\mathrm{corr}}^{100}$. The solid colored symbols represent the median  values in bins of the physical parameter in question. The error bar in the y-direction shows the bin size and in the x-direction shows the error on the median ($\sigma/\sqrt{N}$).The open black symbols represent the median values in bins of radii. In this case, error bars have the same meaning as before, in the inverted directions. Small points represent individual measurements.}
\label{fig:size_sedpars_compilation}
\end{figure*}

The distributions of $r_{e,\mathrm{circ}}$ and $r_{T,\mathrm{corr}}^{100}$ sizes versus M$_{\star}$, SFR, age, $A_V$, and IGM transmission are presented in Figure \ref{fig:size_sedpars_compilation}. Several interesting trends can be observed. 
To quantify the degree of correlation of each parameter with size, we use the \citet{pearson1896} correlation coefficient. In our sample, the null hypothesis (i.e. no correlation)  is excluded at the 3$\sigma$ level if $|r(\mathrm{Pearson})|>0.12$

Our data show a correlation between galaxy size and M$_{\star}$ with larger galaxies having on average higher stellar masses. This supports the correlation reported from other samples \citep[e.g][]{franx2008,ichikawa2012,vanderwel2014,morishita2014} at redshifts $z>2$ and extending it up to $z\simeq4.5$.
This correlation is of similar strength for $r_{T,\mathrm{corr}}^{100}$ as is for $r_e$. We obtain $r(\mathrm{Pearson})=0.09$ for $r_{T,\mathrm{corr}}^{100}$ against $r(\mathrm{Pearson})=0.13$ for $r_e$.  
Interestingly, the most massive galaxies in our sample show a break at stellar masses higher than $\log_{10}(M_\star/M_\odot)>10.5$ in this correlation.  
While, when considering $r_{e,\mathrm{circ}}$, sizes are getting smaller, sizes computed from $r_{T,\mathrm{corr}}^{100}$ increase as we move towards higher stellar masses. This supports in part our analysis that low surface brightness regions in massive galaxies with complex morphology are not taken into account by parametric fitting like done in GALFIT resulting in artificially smaller effective radius (by a factor of $\sim 2$).

We find that the SFR is strongly related to galaxy sizes. For sizes measured with  $r_{T,\mathrm{corr}}^{100}$ we find that larger galaxies have higher star-formation rates, with a positive correlation ($r(\mathrm{Pearson})=0.29$). The correlation is not significant with  $r(\mathrm{Pearson})=0.03$ for $r_{e,\mathrm{circ}}$. Considering the M$_{\star}$--SFR main-sequence of star-forming galaxies in VUDS \citep{tasca2015}, this correlation may be partly reflecting the underlying stellar mass-size relation described above.

Galaxies appear to be older when they are smaller when looking at the relation between age (as defined by the start of star formation, see Thomas et al., submitted) and $r_{T,\mathrm{corr}}^{100}$ ($r(\mathrm{Pearson})=-0.20$). From the hierarchical assembly scenario one would expect that larger galaxies have older stellar populations. On the other hand the less significant relation between $r_{e,\mathrm{circ}}$ and age ($r(\mathrm{Pearson})=-0.10$) may also mean that the derived age is dominated by the age of the last major burst of star formation washing out underlying older stellar populations that may exist in the central brightest clumps from previous starburst episodes.

Galaxies with lower extinction values have smaller median sizes than galaxies with higher dust extinction. This trend is more significant for $r_{T,\mathrm{corr}}^{100}$ ($r(\mathrm{Pearson})=-0.15$) than it is for $r_{e,\mathrm{circ}}$ ($r(\mathrm{Pearson})=0.03$). This hints that extended regions of galaxies have a higher dust content than smaller ones.

\begin{figure*}
\centering
\includegraphics[width=\textwidth]{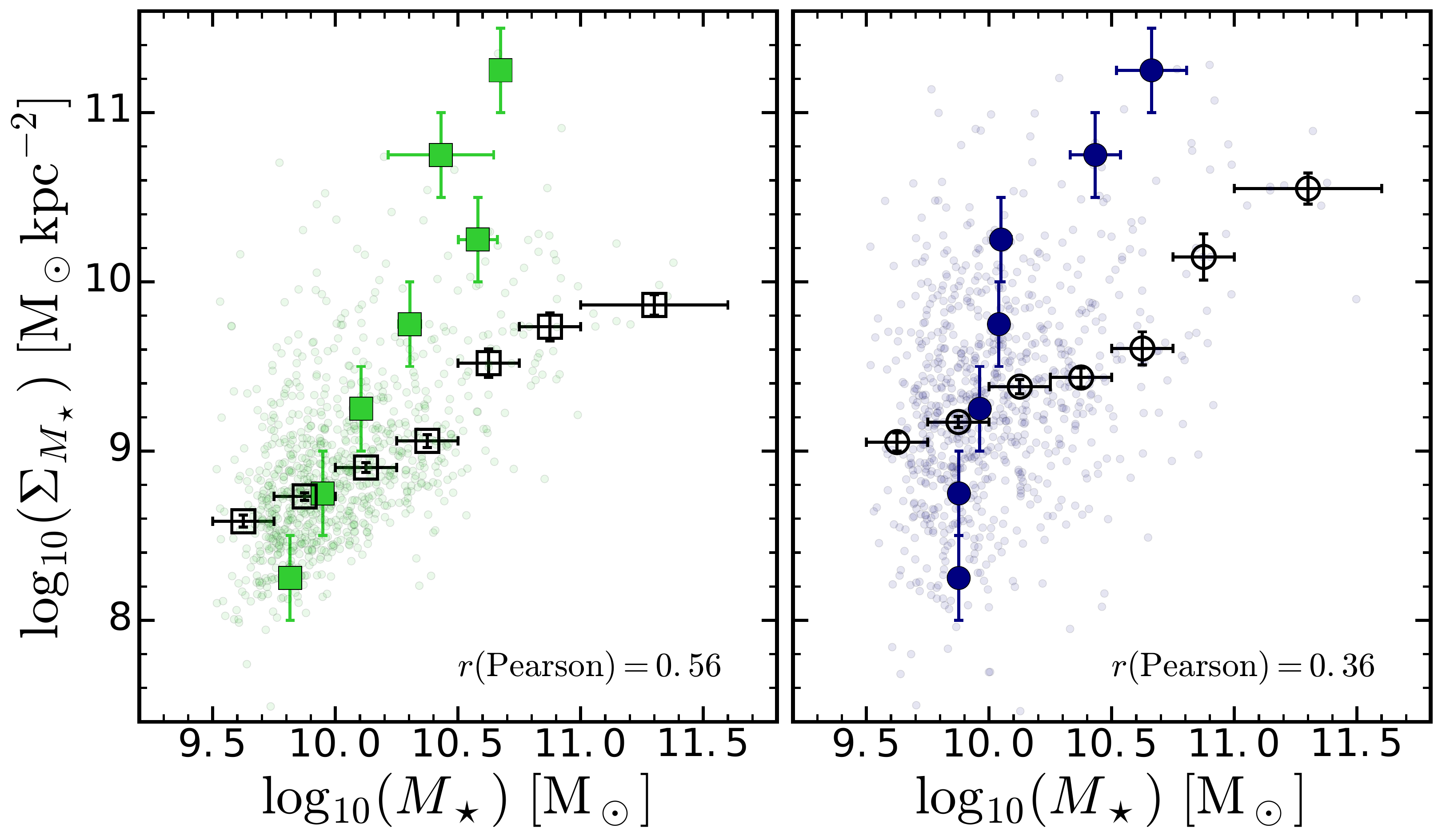}
\caption{Stellar mass surface density as a function of stellar mass. On the right panels we show the results for GALFIT derived $r_{e,\mathrm{circ}}$ and in the left panels we show the results for our value of  $r_{T,\mathrm{corr}}^{100}$. The color and symbols have the same meaning as in figure \ref{fig:size_sedpars_compilation}.}
\label{fig:size_sedpars_mass_density}
\end{figure*}

The projected stellar density computed as the ratio of the stellar mass to the projected area of a galaxy is shown in Figure \ref{fig:size_sedpars_mass_density}. We observe a strong correlation for both  $r_{T,\mathrm{corr}}^{100}$ ($r(\mathrm{Pearson})=0.56$) and $r_{e,\mathrm{circ}}$ ($r(\mathrm{Pearson})=0.36$). Galaxies with higher stellar masses having higher stellar mass densities. 

We do not observe any significant correlation between galaxy sizes and metallicity or IGM transmission for any of our measurements. 

The analysis of spectral properties like the strength of the Lyman$-\alpha$ line as a function of size will be the subject of a forthcoming paper (Ribeiro et al., in preparation).
 

\section{Discussion}\label{sec:discussion}

\begin{figure*}
\centering
\includegraphics[width=\linewidth]{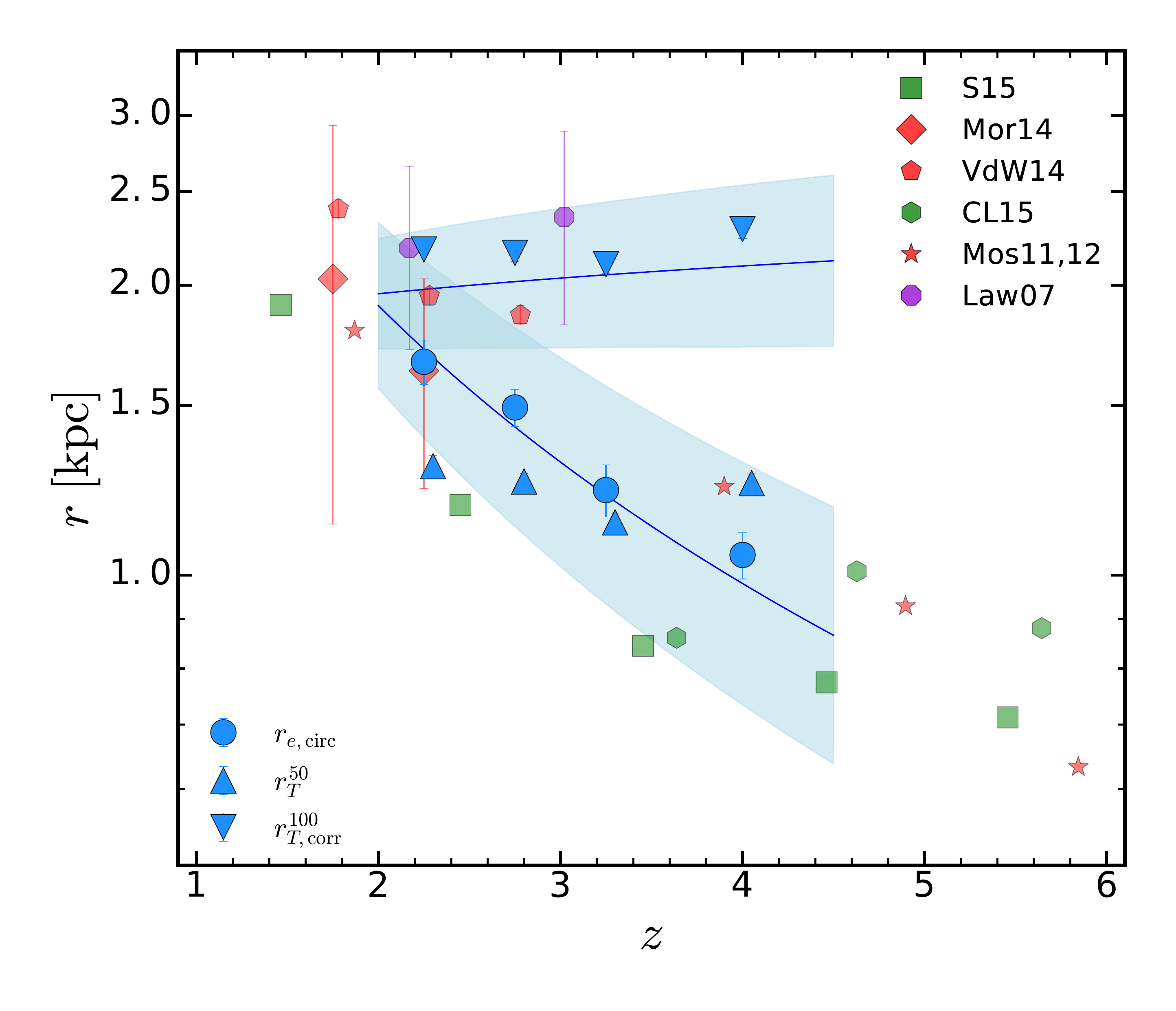}
\caption{A compilation of size measurements at different redshifts for a number of different studies. First, in blue we have the measurements of this paper. Circles are taken from the GALFIT effective radius, and the triangles and inverted triangles are from $r_T^{50}$ and $r_{T,\mathrm{corr}}^{100}$ respectively. The blue solid lines and respective shaded regions show the derived evolution for $r_{e,\mathrm{circ}}$ and  $r_{T,\mathrm{corr}}^{100}$.   Each other color represents a different sample selection (green for luminosity selected samples, red for stellar mass selected samples and purple for a different selection) and each symbol represents a different study. They are: squares \citep{shibuya2015}, $0.3L_*<L<1L_*$; diamonds \citep{morishita2014}, $10.0<\log_{10}(M_\star/M_\odot)<10.5$; pentagons \citep{vanderwel2014} , $10.0<\log_{10}(M_\star/M_\odot)<10.5$; hexagons \citep{curtis-lake2014}, $0.3L_*<L<1L_*$; stars \citep{mosleh2011,mosleh2012}, $9.5<\log_{10}(M_\star/M_\odot)<10.4$; octagons \citep{law2007}, spectroscopic sample, no mass or luminosity selection.}
\label{fig:size_redshift_compilation}
\end{figure*}

We summarize our measurements of the size evolution of star-forming galaxies with $2<z<4.5$ in the VUDS survey, and compare them with the literature in Figure \ref{fig:size_redshift_compilation}. 

In the context of the hierarchical assembly of dark matter halos one can consider that the size of a disk galaxy scales with the virial radius of its host halo. In this scenario, the redshift evolution of galaxy sizes has a slope $\alpha_r=-1$ for halos with fixed mass and a slope of $\alpha_r=-1.5$ for halos with fixed circular velocity \citep[see][and references therein]{ferguson2004,stringer2014}.  We find that when considering the effective radius we obtain a slope $\alpha_r = -1.29\pm0.18$ which lies between the values expected for those two scenarios. However, our computation of the evolution of the total size of each galaxy (measured at the same surface brightness level), leads to a slope $\alpha_r=0.13\pm0.12$ inconsistent with either of these two hypotheses. This likely happens because we are looking at a majority of galaxies that do not resemble disks (i.e they are highly irregular) and, without any constraint on the profile shapes, the size measurements cannot be compared to any of those scenarios. This suggests that to correlate the size growth of galaxies with that of its host dark matter halo, more complex analytical models or more realistic numerical simulations are required to estimate galaxy and halo sizes. 

We find that the effective radius $r_{e,\mathrm{circ}}$ decreases with increasing redshift, indicating effective radii $r_{e,\mathrm{circ}}\sim 2$kpc at $z\sim2$, and a value twice smaller at $z\sim4$
at $r_{e,\mathrm{circ}} \simeq 1$ kpc. The effective radius measured on our galaxies is comparable to other $r_{e,\mathrm{circ}}$ measurements in the literature for galaxies at close or similar redshifts, particularly when comparing to samples with similar stellar masses (Figure  \ref{fig:size_redshift_compilation}). In such samples, as in our own, this leads to the conclusion that galaxies are getting smaller at higher redshifts \citep[e.g.][]{mosleh2011,mosleh2012,vanderwel2014,morishita2014,shibuya2015}. However, we argue that the use of this metric making an a priori assumption of symmetry is not appropriate to constrain the redshift evolution of galaxy sizes when asymmetric and multi-component shapes are common occurrences. 

However when measuring a size based on the total area covered by a galaxy we find that the evolution of size with redshift is strikingly different. This size measured at an isophote with an absolute luminosity $log_{10} (L_{z=2}) = 6.1 L_{\sun}$pc$^2$, after correction for the effects of surface brightness dimming and luminosity evolution, remains approximately constant with redshift with $r_{T,\mathrm{corr}}^{100} \sim 2.2$kpc (Figure \ref{fig:size_redshift_compilation}). In addition, galaxies selected using an evolving stellar mass cut-off $log(M_{\star}) > -0.204(z-4.5) + 9.35$, have a higher concentration index $C_T$ indicating a more peaked light distribution when going to the highest redshifts, with a significant change of $C_T$ by $\sim20$\% over only $\sim2$ Gyr of evolution. 

While it is expected that the effective radius should be smaller than a radius defined from the total size of a galaxy at a much fainter isophote level, the strong observed differential evolution between the two is likely to be resulting from important physical processes working at a cosmic time when galaxies are still early in their assembly process. The fact that the size $r_{T,\mathrm{corr}}^{100}$ remains large at all redshifts indicates that galaxies in their early assembly phase are quite extended with a constant characteristic size, and that the initial collapse that led to star formation seems to have been spread-out in quite a large volume of $\sim 10$kpc$^3$. 

The large dispersion observed around the mean size value may indicate quite a diversity of initial assembly conditions. Indeed we find a large spread of sizes in our sample with galaxies as small as $r_{T,\mathrm{corr}}^{100}=0.3$kpc, and galaxies as large as $r_{T,\mathrm{corr}}^{100}=5.5$kpc. The smallest galaxies could be the signature of the first collapse of a small to medium mass cloud or a more evolved dynamical state later in the life of a galaxy as indicated by larger ages for smaller galaxies, while the most extended ones could indicate on-going early merging of a few small size objects following the hierarchical assembly picture or the result of the multi-component collapse of a massive gas cloud. As the timescale of these processes is currently unknown, we may be witnessing the combined effects of different key physical processes each with different evolution stages.

The evolution in light concentration with redshift that we observe may be understood in a scenario of compact bright clumps merging together to create a more smooth light distribution by $z=2$ \citep[e.g.][]{guo2015}. This could be extended to merger events of compact galaxies redistributing angular momentum and leading to a less concentrated luminosity profile. Inside-out growth with stellar mass building-up in the outskirts of galaxies \citep{wuyts2011,nelson2013,nelson2015,tacchella2015} is also compatible with the evolution of the light concentration and the absence of evolution of total sizes.

\section{Summary}\label{sec:conclusions}

In this paper we study the evolution of the sizes of star-forming galaxies obtained from a detailed analysis of $\sim$1200 galaxies with $2<z<4.5$ and stellar masses $9.5<log(M_{\star})<11.5$ in the VIMOS Ultra Deep Survey. 
We use two different methods to compute sizes: a parametric profile fitting using the GALFIT tool to derive the effective radius $r_{e,\mathrm{circ}}$, and we define and use a non-parametric size measurements based on the area defined as enclosing 100\% of the measured flux of a galaxy above a given surface brightness threshold, $r_T^{100}$. While the former makes a strong hypothesis on the symmetry of a galaxy shape, the latter can be applied on galaxies with irregular and asymmetric morphology. 

Our results on size measurements can be summarized as follows:
\begin{itemize}
\item The total size of a galaxy is observed to remain approximately constant with a radius $r_{T,\mathrm{corr}}^{100} \simeq2.2$ kpc over the redshift range $2<z<4.5$, while the effective radius $r_{e,circ}$ decreases from $\sim$2 kpc at $z\sim2$ to $\sim$1 kpc at $z\sim4.5$. The evolution of galaxy sizes is therefore drastically different when using parametric methods with a symmetry prior or a non-parametric method with no \emph{a priori} assumption of symmetry. This difference between the total radius and the effective radius is more important at the highest redshifts, with a factor $\sim 2$ difference at $z\sim4$, likely due to the increasing fraction of irregularly shaped galaxies.
\item There is a large scatter in galaxy sizes observed at all redshifts. We observe galaxies as large as $\sim11$kpc (total extent) out to $z=4.5$.
\item The projected luminosity density of galaxies is higher at the highest redshifts. Using stacking analysis as well as individual measurements we find that the light concentration of galaxies is $\sim20$\% higher at $z\sim4$ than at $z\sim2$.
\item On average, larger SFGs have higher stellar masses, SFRs anf stellar mass densities.
\item These correlations between size and physical properties depend on the size measurement method. We find a stronger correlation between sizes and star-formation rates for $r_{T,\mathrm{corr}}^{100}$ than for $r_{e,\mathrm{circ}}$ whilst it is of similar strength when considering the stellar mass-size relations.
\end{itemize}

The results presented in this paper on a sample of star-forming galaxies observed 10.5 to 12.5 Gyr ago and 0.5 Gyr after the end of reionization et $z\sim6$ emphasize the diversity of galaxy size properties at a time of intense galaxy assembly. The observation of galaxies with total sizes ranging from less than 1 kpc to $\sim$10 kpc seems to require several different processes driving mass assembly, and may represent different snapshots in the early life history of the dominant population of galaxies at these redshifts. 

More detailed investigations connecting size and morphology analyses to different spectral properties from the UV (e.g. Ly$-\alpha$ emission, $\beta-$slope) to the optical domain are needed to further understand the diversity of sizes in the early phases of galaxy assembly (Ribeiro et al., in preparation).  

Future observations with next generation facilities like the \emph{James Webb Space Telescope} and extremely large telescopes will be necessary to further investigate the nature of large and irregular galaxies at these high redshifts and well into the reionization epoch.


\begin{acknowledgements}
This work is supported by funding from the European Research Council Advanced Grant ERC--2010--AdG--268107--EARLY and by INAF Grants PRIN 2010, PRIN 2012 and PICS 2013. 
This work is based on data products made available at the CESAM data center, Laboratoire d'Astrophysique de Marseille. 
This work partly uses observations obtained with MegaPrime/MegaCam, a joint project of CFHT and CEA/DAPNIA, at the Canada-France-Hawaii Telescope (CFHT) which is operated by the National Research Council (NRC) of Canada, the Institut National des Sciences de l'Univers of the Centre National de la Recherche Scientifique (CNRS) of France, and the University of Hawaii. This work is based in part on data products produced at TERAPIX and the Canadian Astronomy Data Centre as part of the Canada--France--Hawaii Telescope Legacy Survey, a collaborative project of NRC and CNRS.
\end{acknowledgements}


\bibliographystyle{aa}
\bibliography{refs_sizes}

\begin{appendix}
\section{The Concentration parameter}\label{app:concentration}

This concentration index is defined as the ratio of the 80\% to the 20\% curve of growth radii within 1.5 times the \citet{petrosian1976} radius for a parameter $\eta=0.2$:
\begin{equation}
\eta = \frac{\mu(r_p)}{\bar\mu(r<r_p)}
\label{eq:petrosian}
\end{equation}
where $r_p$ is the computed petrosian radius for a given value of $\eta$. With that radius we can compute the flux within circular apertures (or elliptical ones) up to which 20\% and 80\% of the light is contained. But first, it is necessary to define a light center and the axis ratio and position angle so that we can define the elliptical apertures.

To measure this quantities we shall first define the barycenter of the galaxy. The intensity centroid is computed as the mean pixel position, averaged by the flux, within the segmentation map:
\noindent
\begin{eqnarray}
x_{cen} = \frac{1}{I_{tot}}\sum_{i,j} i\times I_{i,j} \label{eq:xcen} \\
y_{cen} =\frac{1}{I_{tot}}\sum_{i,j} j\times I_{i,j} \label{eq:ycen}
\end{eqnarray}
Once one has $\bar x$ and $\bar y$ it is possible to compute the second order moments of the profile as:
 \begin{eqnarray}
\overline{x^2} = \frac{1}{I_{tot}}   \sum_{i,j} i^2 \times I_{i,j} - {\bar x}^2 \label{eq:xx} \\
\overline{y^2} = \frac{1}{I_{tot}}   \sum_{i,j} j^2 \times I_{i,j} - {\bar y}^2   \label{eq:yy}\\
\overline{xy}  = \frac{1}{I_{tot}}    \sum_{i,j} i\times j \times I_{i,j}  - {\bar x}{\bar y}\label{eq:xy}
\end{eqnarray}
From there, it is possible to show that (see \citealt{bertin1996}):
\begin{equation}
\tan 2\theta =2 \frac{\overline{xy}}{\overline{x^2}-\overline{y^2}} ~~\Leftrightarrow~~ \theta = \frac{\mathrm{sign}({\overline{xy})}}{2}\arctan\left(2 \frac{\overline{xy}}{\overline{x^2}-\overline{y^2}} \right)
\end{equation}
Which is then corrected according to the quadrant it is referring in order to span the entire range from -90 to 90 degrees. It is also possible to define directly the semi-major and semi-minor axis of the corresponding ellipse as:
 \begin{eqnarray}
a^2 = \frac{\overline{x^2}+\overline{y^2}}{2} + \sqrt{\left( \frac{\overline{x^2}+\overline{y^2}}{2} \right)^2 + \overline{xy}^2} \label{eq:a2} \\
b^2=\frac{\overline{x^2}+\overline{y^2}}{2} -\sqrt{\left( \frac{\overline{x^2}+\overline{y^2}}{2} \right)^2 + \overline{xy}^2} \label{eq:b2}
\end{eqnarray}
which in turns makes it easier to compute $q=b/a$.

Finally, we can compute the total flux within an elliptical aperture defined by the shape parameters above and a semi-major axis of 1.5$r_p$. From there, we iterate over different apertures with an increasing major-axis to define the radius enclosing 20\% and 80\% of the total flux defined before. Once we have those values it is straightforward to use equation \ref{eq:concentration} to compute the light concentration of the object.

\begin{figure}
\centering
\includegraphics[width=\linewidth]{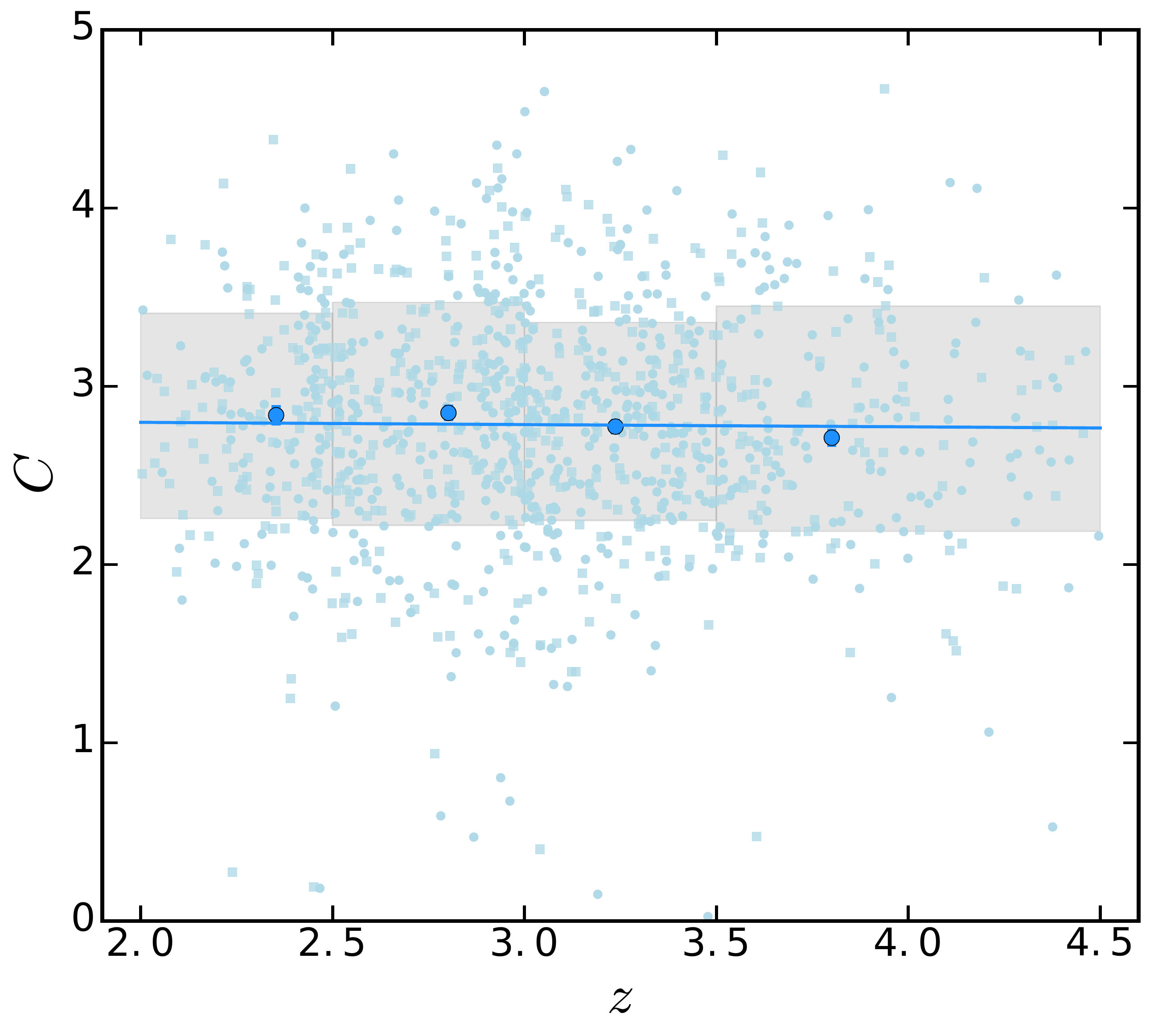}
\caption{The redshift dependence of the standard concentration measurement describe in this section. The symbols and regions have the same meaning as in figure \ref{fig:concentration}.}
\label{fig:standard_CE}
\end{figure}

In figure \ref{fig:standard_CE} it is possible to confirm that for the same sample defined in this paper, the standard light concentration measurement yields different results and trends. This is most likely due to the fact that using the elliptical apertures fails to give a actual insight onto how concentrated is the light inside a galaxy that is highly irregular.

\section{$T_{100}$ in simulated galaxies}\label{app:simulations}

We have used the same set of simulated galaxy images described in section \ref{sec:re_simul} to compute the value of $T_{100}$ and assess the impact of the noise on the retrieved values. To compute the value of the expected total area, we have used the noise free simulated models, convolved with a PSF, and applied the same thresholding method as explained in section \ref{sec:rtot}. To ensure that we would probe different thresholds, for each galaxy, a random redshift was attributed taken from a uniform distribution between the limits we have defined for this paper. This, in practice, translates to mapping the thresholds used in the computation of sizes in real data.

\begin{figure*}
\centering
\includegraphics[width=\linewidth]{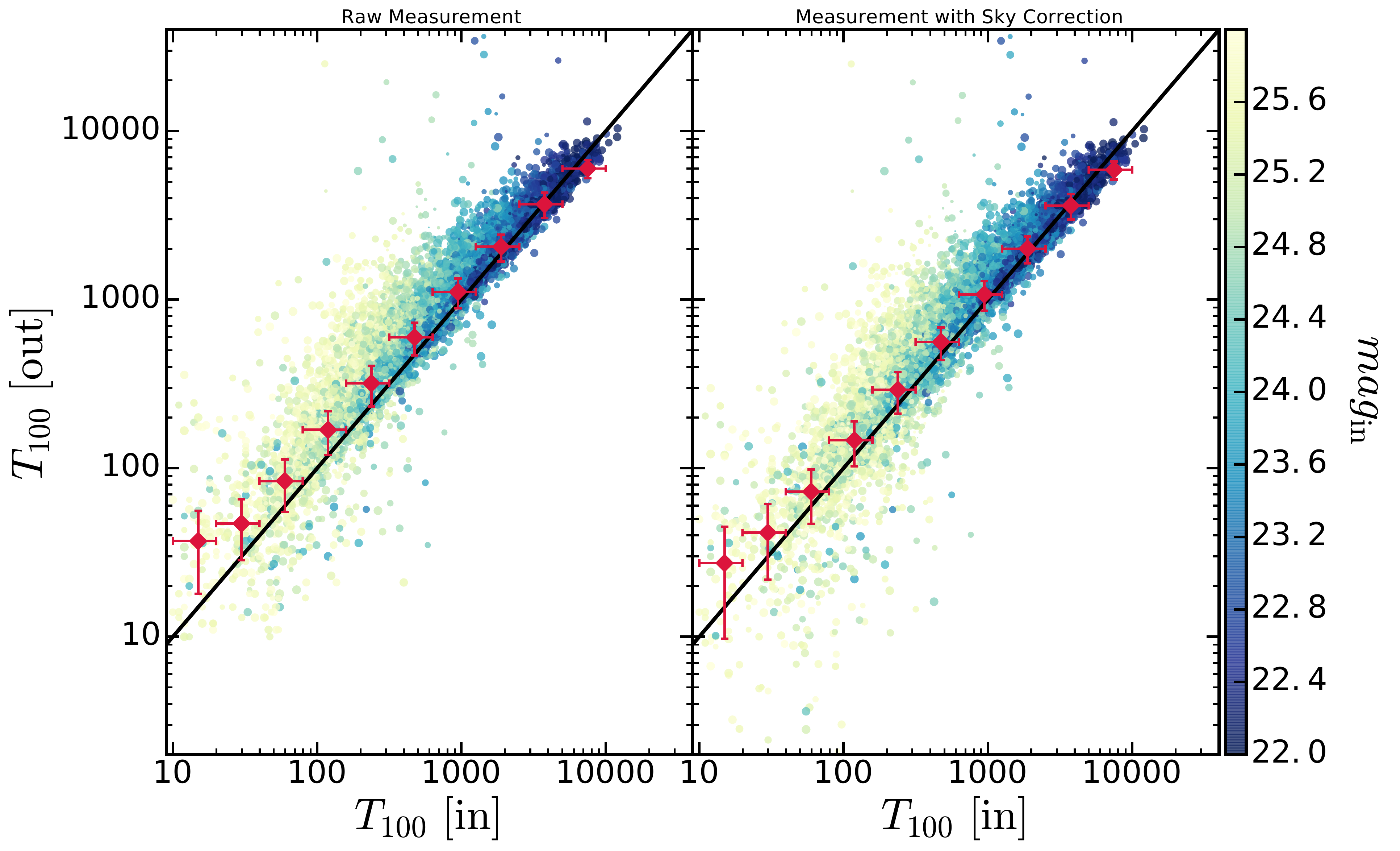}
\caption{Results of the computation of $T_{100}$ for a set of 15 000 galaxies. The points are color coded by their input magnitude and the size of each point is determined by the input $r_e$ of the GALFIT model. The solid red line denotes the one-to-one relation. The left panel refers to the size measurement without any correction and the right panel presents the same results but after applying the same sky correction as described in section \ref{sec:rtot_sky}. The large red points represent the median sizes per bin, the error bar in the x-axis represents the bin width, and in the y-axis represents the median absolute dispersion of the bin.}
\label{fig:simulated_T100}
\end{figure*}

In figure \ref{fig:simulated_T100} we can see the output results from the simulation. As it should be straightforward to understand, brighter galaxies have larger sizes simply because more pixels are found above the threshold that we defined. The scatter that we find increasing for smaller galaxies (fainter magnitudes) is explained by the increasing contamination of the sky connected pixels that are detected close to the galaxy and gain importance as we lower the flux of simulated galaxies which produces an overestimation of the galaxy size. Additionally, since we are dropping galaxy models on top of a real background, with potentially other sources in image area, some large scatter is expected at all magnitudes where sizes are consistently overestimated. It is somehow unexpected the underestimation of galaxy sizes at the largest simulated sources. However, since $\sim90\%$ of the galaxies in our sample have $100<T_{100}<2000$pixels we find it a negligible correction with no impact in our findings.

We have also applied the same correction method for the contamination of sky connected pixels and the results are found in the right panel of figure \ref{fig:simulated_T100}. We can see that it increases somehow the scatter at the lower input sizes and overall approximates the retrieved value from the input one. Being far from perfect, we believe that our sky corrections estimates are good enough for an approximation on the derived galaxy sizes of our sample.

\begin{figure*}
\centering
\includegraphics[width=\linewidth]{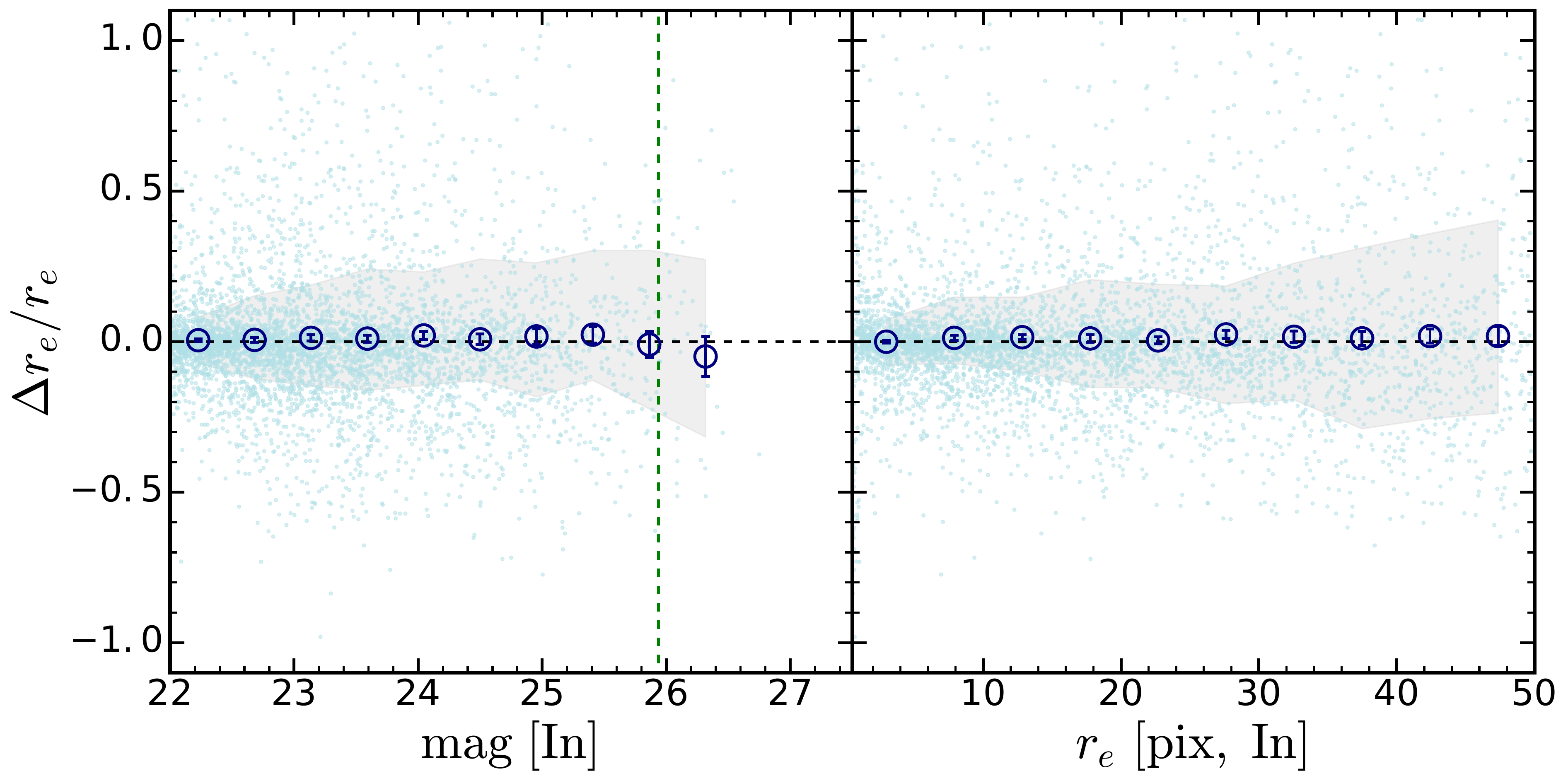}
\caption{Variance of the value of $r_e$ for a set of 15 000 galaxies as a function of input magnitude (left) and effective radii (right). Only galaxies where GALFIT convergence was attained are included in this figure. The large dark blue points represent the median sizes per bin, the error bar in the y-axis is $\sigma/\sqrt{N}$. The shaded region delimits the 16th and 84th percentiles and includes 68\% of the sample of each bin.}
\label{fig:simulated_re}
\end{figure*}

\end{appendix}

\end{document}